\documentclass[3p]{elsarticle}

\usepackage[T1]{fontenc} 
\usepackage{amsmath}
\usepackage{hyperref}

\biboptions{longnamesfirst,semicolon,sort&compress}

\newcommand{\alp}{\alpha^\prime}
\newcommand{\alc}{{\hat\alpha}'}
\newcommand{\alch}{\frac{\hat\alpha^\prime}{2}}

\newcommand{\Tac}{\mathcal{T}}
\newcommand{\Amp}{\mathcal{A}}
\newcommand{\de}{\partial}

\newcommand{\lef}{\langle}
\newcommand{\re}{\rangle}

\newcommand{\pmm}{\frac{p_-}{2}}

\newcommand{\ap}{\alpha^\prime}
\newcommand{\ie}{{\it{i.e.} }}

\newcommand{\be}{\begin{equation}}
\newcommand{\ee}{\end{equation}}
\newcommand{\bea}{\begin{equationarray}}
\newcommand{\eea}{\end{equationarray}}
\newcommand{\bit}{\begin{itemize}}
\newcommand{\eit}{\end{itemize}}

\newcommand{\cA}{\mathcal{A}}

\newcommand{\cD}{\mathcal{D}}
\newcommand{\cE}{\mathcal{E}}
\newcommand{\cF}{\mathcal{F}}
\newcommand{\cG}{\mathcal{G}}
\newcommand{\cH}{\mathcal{H}}
\newcommand{\cI}{\mathcal{I}}

\newcommand{\cK}{\mathcal{K}}
\newcommand{\cL}{\mathcal{L}}
\newcommand{\cM}{\mathcal{M}}

\newcommand{\cO}{\mathcal{O}}

\newcommand{\cR}{\mathcal{R}}
\newcommand{\cS}{\mathcal{S}}
\newcommand{\cT}{\mathcal{T}}
\newcommand{\cU}{\mathcal{U}}

\title{\boldmath On the soft limit of closed string amplitudes \\ with massive states}

\author[rvt]{M.~Bianchi}
\ead{massimo.bianchi@roma2.infn.it}
\author[rvt]{A.~L.~Guerrieri}
\ead{andrea.guerrieri@roma2.infn.it}
\address[rvt]{Dipartimento di Fisica, Università di Roma "Tor Vergata" \\ and Sezione INFN  di Roma II "Tor Vergata" \\ Via della Ricerca Scientifica, 00133 Rome, Italy}

\begin{document}
\begin{abstract}
We extend our analysis of the soft behaviour of string amplitudes with massive insertions to closed strings at tree level (sphere). Relying on our previous results for open strings on the disk and on KLT formulae we check universality of the soft behaviour for gravitons to sub-leading order for superstring amplitudes and show how this gets modified for bosonic strings. At sub-sub-leading order we argue in favour of universality for superstrings on the basis of OPE of the vertex operators and gauge invariance for the soft graviton. The results are illustrated by explicit examples of 4-point amplitudes with one massive insertion in any dimension, including $D=4$, where use of the helicity spinor formalism drastically simplifies the expressions. As a by-product of our analysis we confirm that the `single valued projection' holds for massive amplitudes, too. We briefly comment on the soft behaviour of the anti-symmetric tensor and on loop corrections.
\end{abstract}


\maketitle

\flushbottom

\section{Introduction and motivations}

The connection among `gravitational memory', `soft behaviour' of graviton scattering amplitudes and `BvBMS symmetry' \cite{Bondi:1962px, Barnich:2009se, Barnich:2011mi, Barnich:2011ct, Strominger:2013jfa, He:2014laa} seems to play a crucial in a recently proposed solution to the Information Paradox for Black Holes \cite{Hawking:2015qqa}. While waiting for a refined version of the argument 
, it is natural to ask the fate of the universal `soft' behaviour of graviton scattering amplitudes in a quantum theory of gravity such as closed string theory. The problem has been addressed for tree-level amplitudes with only mass-less gravitons in \cite{MBYHSHCW,Schwab}, relying on KLT formulae and OPE of the vertex operators, and in \cite{DVBernNohl}, relying on gauge invariance. Bosonic amplitudes with tachyons have been investigated to sub-leading order in \cite{DVetal,DiVecchia:2015srk}. 

In gravity theories, when one of the external graviton momenta goes soft \ie $k \rightarrow 0$ with $k=\delta \hat{k}$ with $\hat{k}$ some fixed momentum, not only the leading $\delta^{-1}$ and sub-leading behaviours $\delta^{0}$ \cite{Weinberg1965, GrossJackiw1968}, but also the next-to-subleading or sub-sub-leading behaviour $\delta^{+1}$ is universal \cite{Cachazo:2014fwa}. Calling ${h}^{\mu\nu}_s$ the soft graviton polarisation and $k^\mu_s$ its soft momentum, one has  
\be
\label{softgrav}
\cM_{n}(1,2,\ldots,s,\ldots,n) \approx  \nonumber\ee
\be \sum_{i\neq s} \left[{k_i{\cdot}{h}_s{\cdot}k_{i} \over k_s{\cdot}k_{i}} + { k_i {\cdot} {h}_s{\cdot}J_{i}{\cdot}k_s \over k_s{\cdot}k_{i}} + {k_s{\cdot}J_i{\cdot}{h}_s{\cdot}J_{i}{\cdot}k_s \over 2 k_s{\cdot}k_{i}}\right] \cM_{n-1}(1,2,\ldots\hat{s}\ldots, n) + \cO(\delta^2)
\ee
where $k_i$ and $J_i$ denote the `hard' momenta and angular momentum operators. These results are valid at tree-level and are derived with the understanding that interactions be governed by minimal coupling. 

In theories with closed strings, the conclusions, though quite independent of the number of (non-compact) space-time dimensions, depend on the nature of the higher derivative couplings \cite{MBYHSHCW}. 
$R^3$ terms do not change the universal soft behaviour of minimal coupling, while  $\phi R^2$ do modify even the leading term when $\phi$ is a massless scalar such as the dilaton. This happens in particular in the bosonic string and heterotic string at tree level\footnote{M.~B. would like to thank I.~Antoniadis for stressing the tree level origin of this term in the heterotic string, which only gets generated at one-loop in 4-dim Type II theories with 16 supercharges, such as after compactification on $K3\times T^2$. $R^3$ term is forbidden due to supersymmetry.
} and in the Type II compactifications preserving less than maximal super-symmetry.  

Aim of the present investigation, that may be considered a follow up of~\cite{MBALGopen}, is to show that inclusion of massive external states does not spoil the universal `soft' behaviour~\eqref{softgrav} for Type II theories with maximal susy at tree level. In~\cite{MBALGopen} open string amplitudes with massive external states as well as tachyons have been computed and shown to expose the expected behavior even when non-minimal interactions are considered. Neither $F^3$ terms nor the coupling $\ap \cT F^2$, where $\cT$ is the tachyon, change the universal soft behaviour, based on minimal coupling. On the other hand  $\phi F^2$ terms do modify even the leading term when $\phi$ is a massless scalar.
For color-ordered string amplitudes one gets the same universal behaviour as in YM theories~\cite{Low1954, Casali:2014xpa,He:2014bga,He:2014cra,Lysov:2014csa,Liu:2014vva,Kapec:2014zla,Larkoski:2014bxa,Mohd:2014oja,He:2015zea,Volovich:2015yoa,Campiglia:2015qka,DiVecchia:2015bfa}
\be
\cA_{n}(1,2,\ldots,s,\ldots,n) \approx  \nonumber\ee
\be \left\{ \left[{a_s{\cdot}k_{s+1} \over k_s{\cdot}k_{s+1}} - {a_s{\cdot}k_{s-1} \over k_s{\cdot}k_{s-1}} \right] + \left[{f_s{:}J_{s+1} \over  k_s{\cdot}k_{s+1}} - {f_s{:}J_{s-1} \over  k_s{\cdot}k_{s-1}}\right]\right\} \cA_{n-1}(1,2,\ldots\hat{s}\ldots, n) + \cO(\delta)
\ee
where $a_s$ and $k_s$ denote the soft gluon polarisation and momentum, so that $f^{\mu\nu}_s = k^\mu_s a^\nu_s - k^\mu_s a^\nu_s$ is its linearised field strength, while $k_{s\pm1}$ and $J_{s\pm1}$ denote the `hard' momenta and angular momentum operators of the adjacent insertions.
Relying on \cite{MBALGopen} and on KLT formulae, we presently analyse closed string amplitudes with massive external states. In the bosonic string case we will also consider tachyons as external states. 

Amplitudes with massive external states have been considered earlier on
~\cite{Bianchi:2010dy, Bianchi:2010es, Bianchi:2011se,Black:2011ep}, see also~\cite{Anastasopoulos:2011hj} for the case of `light' string states
and~\cite{DudMou, ChiaIeng, MBAVS,Skliros:2009cs,Skliros:2011si,Feng:2010yx,Feng:2011qc} as well as the review~\cite{AnchLuStiTay} for more phenomenological applications.  
Plan of the paper is as follows. 

In Section~\ref{VSV}, we briefly review KLT formulae relating closed string to open string amplitudes and the `single valued projection' suggested in~\cite{StStSVP,StStTaylor}. Then 
we discuss how to relate the soft limit of closed string amplitudes with an arbitrary number of massive insertions to the soft limit of open string amplitudes with the same number of massive insertions in Section~\ref{Soft}. In Section~\ref{superamps} and~\ref{boseamps} we illustrate our point with explicit examples of 4-point amplitudes with one massive higher spin insertion (or tachyons in the bosonic case). We check the (non) universality of the soft behaviour for bosonic string gravitons in Section~\ref{softexamples} and discuss how to generalise the analysis to the case of anti-symmetric tensors. For the superstrings in $D=4$ we rely on the spinor helicity formalism to simplify our expressions. Our conclusions are presented in Section~\ref{conclusions}.  

\section{From Veneziano to Shapiro-Virasoro according to KLT}
\label{VSV}

Closed-string amplitudes, henceforth denoted by $\cM_n$ to distinguish them from open-string amplitudes, denoted by $\cA_n$, can be efficiently computed relying on KLT formulae \cite{KLT}. At the cost of being pedantic, in order to fix our notation and illustrate the KLT procedure, we start by briefly reviewing some 4-point string amplitudes involving tachyons or massless states.

In going from open to closed strings the mass shell condition becomes $\ap_c (p/2)^2 = (N-1)$ that effectively amounts to the replacement $\ap_o \rightarrow \ap_c/4$\footnote{While the open string spectrum is given by $\ap_{op} M_N^2 = N-1$ with $N = S_{Max}$, the closed string spectrum is given by $\ap_{cl} M_N^2 = 4(N-1) = 
2(N_L+N_R -2)$ due to level matching $N_L=N_R = N = S_{Max}/2$.}. As a result a closed string vertex operator can be expressed as the product of two open-string vertex operators, each carrying half of the total momentum. In formulae
\be V_{cl}(\cH=H\otimes\tilde{H}, p) = V_{op}^L(H, p/2) V_{op}^R(\tilde{H}, p/2),\ee
where $p^2 = m_{\cH}^2 = 4 m_H^2$, and $\cH=H\otimes\tilde{H}$ in general comprises several irreducible representations of the Lorentz group.  

\subsection{Four tachyons: $\cM(\Tac_1,\Tac_2,\Tac_3,\Tac_4)$}

The simplest closed-string amplitude is the Shapiro-Virasoro amplitude $\cM_4(\Tac_1,\Tac_2,\Tac_3,\Tac_4)$ describing the scattering of four tachyons in the closed bosonic string. The tachyon vertex operator is 
\be
V_{\cT}(z,\bar{z})=e^{ipX(z,\bar{z})}= e^{i{p\over 2}X_L(z)}e^{i{p\over 2}X_R(\bar{z})},
\ee
with $\ap_c p^2 = {+}4 = {-}  \ap_c M_{\cT}^2$. 
Up to an overall constant factor,
one finds \cite{Virasoro, Shapiro}
\begin{align}
\cM_4(\Tac_1,\Tac_2,\Tac_3,\Tac_4)&=\pi \int d^2z\,|z|^{\ap_c p_3p_4}|1-z|^{\ap_c p_2p_3} \nonumber\\
&=\pi\frac{\Gamma(1{+}{\ap_c\over 2} p_3p_4)\Gamma(1{+}{\ap_c\over 2} p_2p_3)\Gamma({-}1{-}{\ap_c\over 2} p_3(p_2{+}p_4))}{\Gamma(-{\ap_c\over 2} p_3p_4)\Gamma(-{\ap_c\over 2} p_2p_3)\Gamma(2{+}{\ap_c\over 2} p_3(p_2{+}p_4))},
\end{align}
where use has been made of the integral
\be
\cI(a,n;b,m)=\int d^2z\, |z|^a |1{-}z|^b z^n(1{-}z)^m=\frac{\Gamma(1{+}n{+}\frac{a}{2})\Gamma(1{+}m{+}\frac{b}{2})\Gamma({-}1{-}\frac{a+b}{2})}{\Gamma({-}\frac{a}{2})\Gamma({-}\frac{b}{2})\Gamma(2{+}n{+}m{+}\frac{a+b}{2})}.
\ee
Rewriting the amplitude as a function of the Mandelstam variables $s, t, u$ yields
\be
\cM_4(\Tac_1,\Tac_2,\Tac_3,\Tac_4)=\pi\frac{\Gamma({-}1{-}\frac{\ap_c}{4}s)\Gamma({-}1{-}\frac{\ap_c}{4}t)\Gamma({-}1{-}\frac{\ap_c}{4}u)}{\Gamma(2{+}\frac{\ap_c}{4}s)\Gamma(2{+}\frac{\ap_c}{4}t)\Gamma(2{+}\frac{\ap_c}{4}u)},
\ee
multiplying and dividing by $\Gamma(-1-{\ap_c}t/{4})$, and using the relation
$\Gamma(z)\Gamma(1-z)={\pi}/{\sin\pi{z}}$ produces the KLT relation \cite{KLT}
\be
\cM_4(\Tac_1,\Tac_2,\Tac_3,\Tac_4) = \sin\left(\pi\frac{\alp_c}{4}t\right) \cA_4^L(T_1,T_2,T_3,T_4)\cA_4^R(T_1,T_3,T_2,T_4),
\ee
where $\cA_4(T_1,T_2,T_3,T_4)$ denotes the Veneziano amplitude
\begin{align}
\label{eq:TTTT}
&\cA_4(T_1,T_2,T_3,T_4) = \int_0^1 dx \, x^{{-}\ap_os{-}2}  (1-x)^{{-}\ap_ot{-}2} = {\Gamma\left(-1-\frac{\ap_c}{4}s\right)\Gamma\left(-1-\frac{\ap_c}{4}t\right)\over \Gamma\left(-2-\frac{\ap_c}{4}(s+t)\right)},
\end{align}
where we have used $\ap_{o} \rightarrow \ap_{c}/4$. Henceforth we will set $\ap_{c} = 2$ for convenience.

\subsection{Four massless superstring states: $\cM_4(\cE_1,\cE_2,\cE_3,\cE_4)$}

In type II superstrings the tachyon is projected out. The lowest lying states in the NS-NS sector are massless. 
The massless vertex operator
\be
V_{\cE} = \cE_{\mu\nu} (i\partial X_L^\mu + k\Psi_L \Psi_L^\mu) (i\bar\partial X_R^\nu + k\Psi_R \Psi_R^\mu)
e^{i{k\over 2}X_L(z)}e^{i{k\over 2}X_R(\bar{z})}
\ee
with $k^2 = 0$,  $k^\mu \cE_{\mu\nu} = \cE_{\mu\nu} k^\nu = 0$.
Setting $\cE_{\mu\nu} = \cE_{\nu\mu} = h_{\mu\nu}$, with $\eta^{\mu\nu} h_{\mu\nu} = 0$, describes gravitons, 
$\cE_{\mu\nu} = \cE_{\nu\mu} = \phi_{\mu\nu} = \eta_{\mu\nu} - k_{\mu}\bar{k}_\nu - k_{\nu}\bar{k}_\mu$ describes dilatons, while $\cE_{\mu\nu} = - \cE_{\nu\mu} = b_{\mu\nu}$ describes anti-symmetric tensors (Kalb-Ramond fields). 
For later purposes, it is crucial to observe that gravitons and dilatons are even under L-R exchange, $\Omega = 1$, while Kalb-Ramond fields are odd, $\Omega = - 1$. This implies that amplitudes with an odd number of Kalb-Ramond fields and an arbitrary number of gravitons and dilatons vanish.

The amplitude for 4 massless NS-NS states is well known. The expression is extremely lengthy and 
can be expressed more compactly in terms of the $t_8$ tensor introduced by Brink, Green and Schwarz \cite{BGS}. We refrain from doing so. Using KLT in the $t$-channel, one finds
\be
\cM_4(\cE_1,\cE_2,\cE_3,\cE_4) = \sin\left(\pi\frac{t}{2}\right) \cA^L_4({A}_1,{A}_2,{A}_3,{A}_4) \cA^R_4({A}_1,{A}_3,{A}_2,{A}_4).
\ee
Now writing \cite{MBAVS}
$$
\cA_4^L({A}_1,{A}_2,{A}_3,{A}_4) = { \cF_L^4 \over st} {\Gamma(1-s) \Gamma(1- t)
\over \Gamma(1+u)},
$$ 
with 
$$
\cF_L^4 = \left[(f_1f_2f_3f_4) - {1\over 2}(f_1f_2)(f_3f_4) + {\rm cyclic \: 234}\right]
$$
totally symmetric, and rewriting $\cF_L^4\otimes \cF_R^4 \approx \cR^4 + \ldots$ one can systematically derive  the Type II 4-graviton amplitudes and the related ones for $\phi$'s and (an even number of) $b$'s.

For instance, in $D=4$, $\cF^4$ is only non-vanishing for MHV (Maximally Helicity Violating) configurations \ie $(-,-,+,+)$ or permutations thereof. As a result, $\cF^4 = \langle 1 2\rangle^2 [34]^2$. Similarly, $\cR^4 = \langle 1 2\rangle^4 [34]^4$ for the MHV configurations, \ie (-2,-2,+2,+2). Mixed amplitudes, with gravitons, dilatons and axions arise from combinations with $\cF_L^4\neq \cF_R^4$, for instance, $(-2,0,+2,0) = (-,-,+,+) \otimes (-,+,+,-) =  \langle 1 2\rangle^2 \langle 1 4\rangle^2 [34]^2 [23]^2$ and $(0,0,0,0) = (-,-,+,+) \otimes (+,+,-,-) =  \langle 1 2\rangle^2 \langle 3 4\rangle^2 [34]^2 [12]^2$, while 
$(\pm 2,\pm 2,\pm 2,0) = 0$, $(\pm2,\pm 2,0,0) = 0$, $(\pm 2,0,0,0) = 0$, irrespective of whether the $h=0$ particle is a dilaton or an axion.




For bosonic strings the situation is richer. For open strings the tri-linear coupling is non-minimal. In addition to the standard Yang-Mills term, it contains an $F^3$-term, suppressed by $\ap$. As mentioned in the introduction and discussed in~\cite{MBALGopen}, this does neither spoil universality of the soft behaviour at leading order nor at subleading order, even in the case of massive insertions. For closed bosonic strings, in addition to minimal tri-linear terms (graviton, dilatons and Kalb-Ramond fields), there is a $\phi R^2$ term (suppressed by $\ap$) and an $R^3$-term (suppressed by $(\ap)^2$). As shown in~\cite{MBYHSHCW}, the latter does not spoil the 
universality of the soft behaviour while the former spoils it even at leading order. Barring the distinction between gravitons and dilatons, \ie describing them in a unified fashion with  $\cE_{\mu\nu} = + \cE_{\nu\mu} = 
h_{\mu\nu} + \phi_{\mu\nu}$, one can regain a sort of universality of the soft behaviour as advocated in \cite{DVetal,DiVecchia:2015srk}.
Yet $b_{\mu\nu}$ behaves in a very different way due to its being odd under $\Omega$, as we will see in Section~\ref{boseamps}. 

\subsection{Higher-point amplitudes}

Closed-string amplitudes with massive insertions look extremely cumbersome and not very illuminating in $D=10$, even at tree level (sphere). In $D=4$, using the spinor helicity basis, formulae look more tractable. A possible strategy for systematic computations is to first use KLT relations in order to express closed-string amplitudes in terms of open-string amplitudes, and then compute open-string amplitudes for massive states by multiple factorizations of amplitudes with only massless insertions on massive poles in two-particle channels as in~\cite{MBALGopen}. 

KLT relations incorporate the intrinsic non-planarity of closed-string amplitudes and rely on the monodromy properties of (colour-ordered) open string amplitudes \cite{KLT}. The basic idea is to parameterize the closed-string insertion points as $z_i = x_i +i y_i$ 
and notice that the integrand is an analytic function of the $y_i$ viewed as complex variables with branch points at $\pm i (x_i-x_j)$. One can then deform the integration contour from ${\rm Im}y_i = 0$ to ${\rm Re}y_i = 0$ so much so that $z_i$ and $\bar{z}_i= x_i -i y_i $ become two independent real variables $\xi_i$ and $\eta_i$ that one can integrate over with Jacobian $\partial(x_i, y_j)/ \partial(\xi_i, \eta_j) = (i/2)^N$. The correct monodromy around the branch points of the integrand  (Koba-Nielsen factor, in units $\alpha'_{c} =2$)
$$
\prod_{i>j} (z_i - z_j)^{k_ik_j +n_{ij}} (\bar{z}_i - \bar{z}_j)^{k_ik_j+\bar{n}_{ij}}  \rightarrow
\Phi(\sigma_\xi, \sigma_\eta) \prod_{i>j} (\xi_i - \xi_j)^{k_ik_j +n_{ij}} (\eta_i - \eta_j)^{k_ik_j+\bar{n}_{ij}}
$$
with $n_{ij}$ and $\bar{n}_{ij}$ integer, is accounted for by the phase factor 
$$
\Phi(\sigma_\xi, \sigma_\eta) = \prod_{i>j} \exp\{i \pi k_i k_j \theta[-(\xi_i - \xi_j)(\eta_i - \eta_j)]\}
$$
that only depends on the orderings $\sigma_\xi$ and $\sigma_\eta$ but not on the variables  $\xi$'s and $\eta$'s themselves. The integrations decouples and can be performed explicitly.
In particular, using $SL(2)$ to fix 3 $\xi$'s, there remain $(n-3)!$ orderings of the $\xi$'s. For each of them, the independent choices of the contours in $\eta$ that give a non-vanishing result  give in fact all the same result.  All in all there are $(n-3)![{1\over2} (n-3)!]^2$ terms for $n$ odd or 
$(n-3)![{1\over2} (n-4)!] [{1\over2} (n-2)!]$ for $n$ even \cite{KLT}.
In particular, for $n=3,4$ there is only one term\footnote{Neglecting overall constants.}
$$
 \cM_3(123) = \cA^L_3(123) \cA^R_3(123) 
 $$
 and
 $$ \cM_4(1234) = \sin(\pi k_1k_2)\,  \cA^L_4(1[2]34) \cA^R_4(\underline{2}134). $$
For $n=5$ one has two terms 
\begin{align}
\cM_5(12345) &= \sin(\pi k_1k_2)  \sin(\pi k_3k_4) \cA^L_5(1[23]45) \cA^R_5(\underline{2}14\underline{3}5)  \nonumber\\
&+\sin(\pi k_1k_3)  \sin(\pi k_2k_4) \cA^L_5(1[32]45) \cA^R_5(\underline{3}14\underline{2}5),
\end{align}
while for $n=6$ one has twelve terms
\begin{align}
\cM_6(123456) &= \sin(\pi k_1k_2)  \sin(\pi k_4k_5) \cA^L_6(1[234]56) \nonumber\\
&\left\{ \sin(\pi k_3k_5) \cA^R_6(\underline{2}15\underline{34}6) + \sin(\pi k_3(k_4+k_5)) \cA^R_6(\underline{2}15\underline{43}6)\right\} 
+ {\rm Perm} [234]\nonumber\\
&= \sin(\pi k_1k_2)  \sin(\pi k_4k_5) \cA^L_6(1[234]56) \nonumber\\
&\left\{ \sin(\pi k_1k_3) \cA^R_6(\underline{23}15\underline{4}6) + \sin(\pi k_3(k_1+k_2)) \cA^R_6(\underline{32}15\underline{4}6)\right\} 
+ {\rm Perm} [234].
\end{align}
In general, one has \cite{Vanhoveetc} 
\begin{align}
&\cM_n (1,2,\ldots,n) = \cA_n^L(1,[2,\ldots,n{-}2], n{-}1, n) \nonumber\\
&\sum_{\{i\},\{j\}} f(i_1,\ldots ,i_{\lfloor n/2\rfloor - 1}) 
\tilde{f}(j_1,\ldots ,j_{\lfloor n/2\rfloor{-} 2})\cA_n^R(\{i\},1,n{-}1,\{j\},n) + {\rm Perm} [2,\ldots ,n{-}2],
\end{align}
where $\{i\}\in {\rm Perm} [2,\ldots ,\lfloor n/2\rfloor]$, $\{j\}\in {\rm Perm} [\lfloor n/2\rfloor{+}1,\ldots ,n{-}2]$, with $\lfloor n/2\rfloor = (n{-}1)/2$ for $n$ odd, and $\lfloor n/2\rfloor = n/2 {-} 1$ for $n$ even,
while the relevant momentum kernels read \cite{Vanhoveetc}
\begin{align}
&f(i_1,\ldots i_{m}) = \sin(\pi{s_{1i_m}}) \prod_{k=1}^{m-1} \sin\left(\pi\left({s_{1i_k}} +\sum_{l=k+1}^m 
\hat{s}_{i_ki_l}\right)\nonumber\right),\\
&\tilde{f}(j_1,\ldots j_m) = \sin(\pi{s_{j_1n{-}1}}) \prod_{k=2}^{m} \sin\left(\pi\left(s_{j_kn{-}1} + \sum_{l=1}^{k-1}  \hat{s}_{j_lj_k}\right)\right),
\end{align}
where $\hat{s}_{ij} =s_{ij} = k_ik_j$, if $i>j$, and zero otherwise.
Let us observe once again that KLT formulae are valid for all kinds of closed strings, Type II, Heterotic and Bosonic, at tree level and for any kind of insertions: tachyonic, mass-less or massive.
 
Similar formulae relating string amplitudes with only massless insertions to SYM amplitudes~\cite{MSST,Mafra:2011nw}, see also \cite{BarreiroMedina}, have been derived for open superstrings, whose validity we have given further support in \cite{MBALGopen}. MSST formulae read
\be 
\cA_{ST}(1,\rho [2,\ldots,n{-}2],n{-}1,n) = \sum_{\sigma\in \cS_{n{-}3}} \cF_n[\rho|\sigma] \cA_{YM}(1,\sigma[2,\ldots,n{-}2],n{-}1,n)
\ee
where the $(n{-}3)!\times (n{-}3)!$ dimensional matrices of generalised Euler integrals read 
\be
\cF_n[\rho|\sigma] = (-1)^{n{-}3}(\sqrt{\alp})^{n{-}4} \int_{D(\rho)} \prod_{l=2}^{n{-}2} dz_l \prod_{i<j} z_{ij}^{2\ap k_ik_j} 
\prod_{k=2}^{n-2} \sum_{m=1}^{k-1} { 2\alp k_mk_k \over z_{mk}}
\ee
with integration domain $D(\rho) = \{ 0=z_1<\rho(z_2) <\ldots<\rho(z_{n{-}2})< z_{n{-}1}=1 < z_n=\infty\}$.

Following the strategy outlined above, one can now combine the virtues of KLT and of MSST.
For instance, at 5-points a closed (super)string amplitude with $n$ massless and $m=5{-}n$ massive states, according to KLT, reads 
\be 
\cM_{n,5-n}(12345) = \sin(\pi s_{12}/2) \sin(\pi s_{34}/2)\cA^L_{n,5-n}(1[23]45)\cA^R_{n,5-n}(\underline{2}14\underline{3}5) \nonumber \ee
\be
\quad + \sin(\pi s_{13}/2) \sin(\pi s_{24}/2)\cA^L_{n,5-n}(1[32]45)\cA^R_{n,5-n}(\underline{3}14\underline{2}5) 
\ee
In turn, the open string amplitude $\cA^{L/R}_{n,5{-}n}(12345)$ can be computed factorizing 
$\cA^{L/R}_{10{-}n,0}(1{.}{.}{.} 5)$ on $5-n$ massive poles in two-particle channels. The massless amplitude $\cA^{L/R}_{10-n,0}(1{.}{.}{.} 5)$ can be expressed in terms of $\cA^{SYM}_{10-n}(1{.}{.}{.} 5)$ thanks to MSST formula.
The generalisation, relating $\cM_{n,m}$ with arbitrary $n$ and $m$ to $\cA^{L/R}_{n,m}$ and the latter to $\cA^{L/R}_{n+2m,0}$ and finally to $\cA^{SYM}_{n+2m}$ is straightforward, but more and more cumbersome as the number of particles increases.  

\subsection{From open to closed via `single-valued projection'}
\label{single-valued projection}

Although we will not fully exploit it in the following, an alternative and elegant expression of closed superstring amplitudes with massless insertions only in terms of SYM amplitudes at tree level has been found in~\cite{StStSVP,StStTaylor} that exposes the cancellation of various MZV (Multiple Zeta Values) including rational multiples of $\zeta_{2n}$ in the $\ap$ expansion.  

The `single-valued projection' formula reads\footnote{Notice the exchange of $n$ and $n{-}1$ in ${\widetilde\cA}^{YM}_n(1,2_\rho, 3_\rho, ...(n{-}2)_\rho, n, n{-}1)$.}
\begin{align}
{\cM}_n = \sum_{\rho,\sigma,\tau \in S_{n-3}} {\widetilde\cA}^{YM}_n(1,2_\rho, 3_\rho, ...(n{-}2)_\rho, n, n{-}1) \cS_0[2_\rho, 3_\rho, ...(n{-}2)_\rho| 2_\sigma, 3_\sigma, ...(n{-}2)_\sigma] \nonumber \\ 
\cG_n[\sigma|\tau] 
\cA^{YM}_n(1,2_\tau, 3_\tau, ...(n{-}2)_\tau, n{-}1, n) 
\end{align}
where $\cS_0[\rho|\sigma] = \cS_{KLT}[\rho|\sigma]\vert_{(\ap)^{n-3}}$ 
\be 
\cS_0[\rho(2,\ldots,n{-}2)|\sigma(2,\ldots,n{-}2)] = \prod_{i=2}^{n{-}2} \left(-k_1k_{\rho(i)} - 
\sum_{j=2}^{i{-}1} \theta_{\sigma}(\rho(i),\rho(j)) k_{\rho(i)}k_{\rho(j)} \right)
\ee
with $ \theta_{\sigma}(\rho(i),\rho(j)) =1$ if the ordering of $(\rho(i),\rho(j))$ is equal to the ordering of $(\sigma(i),\sigma(j))$ and zero otherwise. $\cS_0[\rho|\sigma]$
is the `super-gravity' limit of the KLT momentum kernel such that $\sin(\pi\ap k_ik_j/2) \rightarrow  \pi\ap k_ik_j/2$ and the 
$(n-3)!{\times}(n-3)!$ matrix  $\cG_n[\rho|\sigma]$ is given by the `single-valued projection'
\begin{align}
\cG_n[\sigma|\tau] &= 1 {+}\zeta_3 M_3 {+} \zeta_5 M_5 {+} {1\over 2}  \zeta^2_3 M_3M_3 {+} 2\zeta_7 M_7 {+} \ldots = {\rm sv} \{\cF_n[\sigma|\tau] \} \nonumber \\
&= { \rm sv} \left\{1{{+}}\zeta_2P_2{{+}}\zeta_3M_3{{+}}\zeta_2^2P_4{{+}}\zeta_5M_5{{+}}\zeta_2\zeta_3 P_2M_3{{+}}\zeta_2^3 P_6 \right. \nonumber\\
& \left. {{+}}{1\over 2}\zeta^2_3M_3M_3{{+}}2\zeta_7M_7{{+}}\zeta_2\zeta_5 P_2M_5{{+}}\zeta^2_2\zeta_3P_4M_3{{+}}\ldots \right\}
\end{align}
of the $(n-3)!{\times}(n-3)!$  matrix $\cF_n[\rho|\sigma]$ that appear in MSST formula.
Not only all $P_{2n}$ matrices drop but also higher depth MZV's do as a result of properties of the $M_{2k+1}$ matrices. 

\section{Soft limit from open to closed}
\label{Soft}

When considering the soft behaviour of string amplitudes one may expect corrections from standard field 
theory results due to the non-minimal higher-derivative terms in the coupling among mass-less states as well as with massive states. For open strings we have checked that this higher-derivative couplings coded in the OPE of the vertex operators do not spoil universality of the soft behaviour at leading and sub-leading order. 
For completeness, let us now recall the argument \cite{MBYHSHCW, MBALGopen,Schwab}. The OPE of a massless vector boson vertex operator (in the $q=0$ super-ghost picture) and a massive higher spin vertex opeartor (in the $q=-1$ super-ghost picture) reads 
\be
V_A(a_s, k_s) V_M(H_{s\pm 1}, p_{s\pm 1})  \approx {1 \over 2 k_s p_{s\pm 1}} V_{M'}(H' [a_s,H_{s\pm 1}, k_s, p_{s\pm 1}], k_s + p_{s\pm 1}) + ...
\ee
where $M'$ denotes any state at the same mass level as the state $M$. 
For totally symmetric tensors of the first Regge trajectory at level $N=\ell -1$ one has
\begin{align}
&\cA_3({A}_1,H_{2,\ell},H_{3,\ell}) = \nonumber\\
&a_1 p_{23} H_2^{\mu_1\ldots\mu_\ell} H_{3,\mu_1\ldots\mu_\ell} + a_{1,\mu}  H_2^{\mu\mu_2\ldots\mu_\ell} p^\nu_{12}H_{3,\nu\mu_2\ldots\mu_\ell} + p_{31,\mu} H_2^{\mu\mu_2\ldots\mu_\ell} a^\nu_{1}H_{3,\nu\mu_2\ldots\mu_\ell} 
+ \cO(\ap p^2) ].
\end{align}
The leading term  encodes minimal coupling. The sub-leading term is fixed by gauge invariance so that, barring some subtleties, to be dealt with momentarily, one gets 
\begin{align}
&\cA_{n+1,m}(1,\ldots {s} \ldots,n+m+1) \nonumber\\
&\approx \pm \left\{ {a_s{\cdot}p_{s{+}1} \over 2k_s{\cdot}p_{s{+}1}} - 
\ell {k_s{\cdot}H^{\ldots}_{s{+}1} \over 2k_s{\cdot}p_{s{+}1}} a_s {\cdot} {\partial \over \partial H^{\ldots}_{s{+}1}}
 + {a_s{\cdot}p_{s{+}1} \over 2k_s{\cdot}p_{s{+}1}} k_s {\cdot} {\partial \over \partial p_{s{+}1}} + \ell {a_s{\cdot}H^{\ldots}_{s{+}1} \over 2k_s{\cdot}p_{s{+}1}} k_s {\cdot} {\partial \over \partial H^{\ldots}_{s{+}1}}\right\}\nonumber\\
&\qquad\mp {k_s{\cdot}p_{s{+}1} \over 2k_s{\cdot}p_{s{+}1}} a_s {\cdot} {\partial \over \partial p_{s{+}1}} \cA_{n,m}(1,\ldots \hat{s} \ldots,n+m+1) + \ldots,
 \end{align}
for an amplitude with $n$ massless and $m$ massive states.

Before generalising the above argument to the closed string case, let us deal with a couple of subtleties: the higher derivative terms in the tri-linear coupling $A$-$H$-$H$ and the possible non-diagonal couplings $A$-$H$-$H'$ that would spoil universality. First, higher derivative corrections to minimal coupling can only affect the sub-leading term that is fixed by gauge invariance wrt the soft gluon \cite{DVBernNohl}.  Second, for open superstrings already at the first massive level one finds two kinds of particles in the Neveu-Schwarz sector: 
$C_{\mu\nu\rho}$ and $H_{\mu\nu}$.
In addition to the `diagonal' couplings $V${-}$C${-}$C$ and $V${-}$H${-}$H$ (and SUSY related) one should consider the mixed coupling $V${-}$H${-}$C$ $\approx \ap M p_{31}{\cdot}H_2{\cdot}C_3{:}[a_1p_{12}]$ that exposes the singular soft factor $1/kp$ since $M_C = M_H$ but gets suppressed by an extra power of the soft momentum in the numerator. Lacking the leading $\delta^{-1}$ term that fixes also the sub-leading $\delta^{0}$ term, thanks to gauge invariance, this kind of higher derivative non-diagonal couplings can at most affect the sub-sub-leading $\delta^{+1}$ (and higher) terms which are not expected to be universal. 

Relying on KLT, similar arguments were advocated to warrant universality of closed super-string amplitudes to leading, sub-leading and sub-sub-leading order \cite{MBYHSHCW, Schwab}. 
Indeed, the relevant OPE's of closed string vertex operators are simply the L+R combinations of the ones shown above for open strings. This implies that the leading behavior is completely fixed by the trilinear coupling. If this is minimal as for the superstrings one gets a universal behavior if it is not, as for the bosonic and heterotic strings one expects non universality or some sort of generalization thereof \cite{DVetal}. The additional ingredients are two. First, KLT formulae produce amplitudes with non-planar duality, with the soft graviton that can attach to each of the `hard' (massless or massive) legs. Second, not only the sub-leading but also the sub-sub-leading term is fixed by gauge invariance of the soft graviton \cite{DVBernNohl}. We would like to stress that this is true also for amplitudes with massive insertions as we will now sketch  and check with explicit examples later on.
Given universality of the soft behavior of all open string amplitudes for granted \cite{MBALGopen} one schematically has 
\begin{align}
\cM_{n{+}1} &{=} \sum_I \prod_I  \sin(\pi k k')_I \cA^L_{n{+}1}(...) \cA^R_{n{+}1}(...) \nonumber\\
&{\approx} \sum_I \prod_I  \sin(\pi k k')_I  (\cS^{(0)}_L {+} \cS^{(1)}_L {+} ...)\cA^L_n(...) (\cS^{(0)}_R {+} \cS^{(1)}_R {+} ...) \cA^R_n(...)\nonumber\\
&{=} (\cS^{(0)}_c {+}  \cS^{(1)}_c {+} \cS^{(2)}_c {+} ...) \sum_I \prod_I  \sin(\pi k k')_I  \cA^L_n(...) \cA^R_n(...) 
\end{align}
One can easily check that  $\cS^{(0)}_{grav} = \cS^{(0)}_L \cS^{(0)}_R$ using momentum conservation,  similarly
$\cS^{(1)}_{grav} = \cS^{(1)}_L \cS^{(0)}_R + \cS^{(0)}_L \cS^{(1)}_R$. Finally  the sub-sub-leading
$\cS^{(2)}_{grav}  = \cS^{(1)}_L \cS^{(1)}_R + \cS^{(0)}_L \cS^{(2)}_R + \cS^{(2)}_L \cS^{(0)}_R$ to be checked on a case by case basis since  $\cS^{(2)}_{L/R}$ is not universal, but conspires with the permutation to give something universal. We will limit ourselves to check cancellation of $\pi^2 = 6 \zeta_2$ and similar terms that are forbidden by the single-valued projection \cite{StStSVP,StStTaylor}. 
At the cost of being pedantic we would like to reiterate that once the leading term is fixed and universal then sub-leading and sub-sub-leading terms follow thanks to gauge invariance of the soft graviton.                                     

\subsection{4-point amplitudes with massive states}
Let us consider first 4-point amplitudes.
We already know that 
\be
\cM_4(1234) = \sin(\pi \,p_1k_4) \cA^L_4(1234) \cA^R_4(1324),
\ee
allowing for a time-like $p_1$, while we assume $k_4$ to be light-like and `soft' with `polarisation' $\cE=a_L\otimes a_R$. 
From KLT we also know that 
\be
\cA_4(1234) = \cS^{3{-}1}_{4} \cA_3(123)\qquad \text{and}
\qquad \cA_4(1324) = \cS^{2{-}1}_4  \cA_3(123),
\label{ALAR}
\ee
where
\be
\cS_i^{j-l} = \cS_{i(0)}^{j-l} + \cS_{i(1)}^{j-l} + \left(\cS_{i(2)}^{j-l}-\zeta_2 \, k_sp_j\,k_sp_l\,\cS_{i(0)}^{j-l}\right),
\label{Sop}
\ee
with universal
\be
\cS_{i(0)}^{j-l}   = {a_ik_j \over k_ik_j} - {a_ik_l \over k_ik_l} \qquad  
{\rm and}  \qquad \cS_{i(1)}^{j-l} = {f_iJ_j \over k_ik_j} - {f_iJ_l \over k_ik_l}
\ee
while 
\be
\cS_{i(2)}^{j-l} = {f_iW_j k_i\over k_ik_j} - {f_iW_l k_i \over k_ik_l}
\ee
is not universal. In $D=4$ there is only one gauge invariant non vanishing derivative of $f$, \ie 
$u_\alpha \bar{u}_{\dot\alpha} (u_\beta u_\gamma)$ or $u_\alpha \bar{u}_{\dot\alpha} (\bar{u}_{\dot\beta} \bar{u}_{\dot\gamma})$
and $W$ should reflect this structure (pretty much as $J$ parallels $f$ itself). 
The obvious guess is a mixed-symmetry tensor (`hook' Yang tableau) 
$W_{[\lambda(\mu]\nu)} = p_\lambda {\partial^2/\partial p^\mu \partial p^\nu} \pm \ldots$.
Moreover, it is worth to notice that the factor $\zeta_2 = \pi^2/6$ in Eq.~\eqref{Sop} comes from the expansion of the beta function appearing in the open string disk amplitudes 
with four external legs.

Combining the two amplitudes in Eq.~\eqref{ALAR}, and using $\cM_3(123) = \cA^L_3(123) \cA^R_3(123)$ (up to an overall factor) as well as 
$\sin(\pi p_1k_4) = \pi p_1k_4 - \pi^3 (p_1k_4)^3/6 + \ldots$, we get
\be
\cM_4(1234) \approx [\pi p_1k_4 - \pi^3 (p_1k_4)^3/6] \,\cS_4^{3{-}1} \cS_4^{2-1} \cM_3(123) 
\ee
Expanding at leading order yields
\be
\cM_4(1234)\approx \left\{ {p_1\cE_4p_1 \over p_1k_4} + {p_3\cE_4p_2 \, p_1k_4 \over p_2k_4\; p_3k_4} - {p_3\cE_4p_1 \over p_3k_4} - {p_1\cE_4p_2 \over p_2k_4} \right\} \cM_3(123),
\ee
and relying on momentum conservation, and on the standard trick
\be
{p_1k_4 \over p_2k_4\; p_3k_4} = - {1 \over p_2k_4} - {1\over p_3k_4},
\ee
we get
\be
\cM_4(1234)\approx \left\{ {p_1\cE_4p_1 \over p_1k_4} + {p_2\cE_4p_2 \over p_2k_4} +  {p_3\cE_4p_3 \over p_3k_4} \right\} \cM_3(123).
\ee
Only the symmetric (not necessarily trace-less) part contributes, thus exposing the violation of the principle of equivalence in presence of a massless dilaton.  

At sub-leading order one has
\begin{align}
&\cM_4(1234)\nonumber\\
&\quad{\approx} p_1k_4 \left\{ \left[{p_3a^L_4 \over p_3k_4} {-}  {p_1a^L_4 \over p_1k_4}\right]\left[{f^R_4 J^R_2\over p_2k_4} {-}  {f^R_4 J^R_1\over p_1k_4} \right] 
{+}  \left[{J^L_3 f^L_4 \over p_3k_4} {-}  {J_1^Lf^L_4 \over p_1k_4} \right] 
\left[{a^R_4p_2\over p_2k_4} {-}  {a^R_4 p_1\over p_1k_4}\right] \right\} \cM_3(123).
\label{AAA}
\end{align}
Expanding and combining the terms appearing in Eq.~\eqref{AAA}, one gets for the pole in $p_1k_4$
\be
p_1a^L_4 \, f^R_4 J^R_1 + J_1^Lf^L_4 \, a^R_4 p_1 = p_1 \cE^{\pm}_4 (J_1^L\pm J_1^R),
\ee
depending on the `symmetry' of $\cE_4$.
Moreover, for the  pole in $p_2k_4$ one gets 
\be
- (p_1{+}p_3) a^L_4 \, f^R_4 J^R_2 - (J_1^L+J_3^L) f^L_4 \, a^R_4 p_2  = p_2 \cE^{\pm}_4 (J_2^L\pm J_2^R),
\ee 
where  in the last step we used the angular momentum conservation $(J_1^L+J_2^L +J_3^L) \cA_3(123) =0$. 
For  the  pole in $p_3k_4$ one gets the same result {\it mutatis mutandis}.

At sub-sub-leading order one has many terms
\begin{align}
&\cM_4(1234)\nonumber\\
&\approx p_1k_4 \left\{ \left[{p_3a^L_4 \over p_3k_4} {-}  {p_1a^L_4 \over p_1k_4}\right] \left[{u^R_4 W^R_2\over p_2k_4}{ -}  {u^R_4 W^R_1\over p_1k_4} \right] 
{+} \left[{W^L_3 u^L_4 \over p_3k_4} {-}  {W_1^Lu^L_4 \over p_1k_4} \right] \left[{a^R_4p_2\over p_2k_4} {-}  {a^R_4 p_1\over p_1k_4}\right]
\right.
\nonumber \\
&{+} \left.  \left[{J^L_3 f^L_4 \over p_3k_4} {-}  {J_1^Lf^L_4 \over p_1k_4} \right] \left[{f^R_4 J^R_2\over p_2k_4} {-}  {f^R_4 J^R_1\over p_1k_4} \right] 
{-}  \frac{\pi^2}{6} p_1k_4 \left[{p_3a^L_4 \over p_3k_4} {-}  {p_1a^L_4 \over p_1k_4}\right]
\left[{a^R_4p_2\over p_2k_4} {-}  {a^R_4 p_1\over p_1k_4}\right] \right.\nonumber\\
& {-}\left. \zeta_2(k_4p_3+k_4p_2)\left[{p_3a^L_4 \over p_3k_4} {-}  {p_1a^L_4 \over p_1k_4}\right]\left[{a^R_4p_2\over p_2k_4} {-}  {a^R_4 p_1\over p_1k_4}\right]
 \right\} \cM_3(123),
\end{align}
where $u^{L/R}_4 = k_4 k_4 a^{L/R}_4$ and $W^{L/R} = a^{L/R}_4 \,\partial^2/\partial k_4\partial k_4$ properly (anti)symmetrized but not universal (for open strings). 

After lengthy manipulations one reproduces
\be
\cM_4(1234)\approx \left\{ {k_4 J_1\cE_4J_1 k_4 \over p_1k_4} + {k_4 J_2\cE_4J_2 k_4 \over p_2k_4} +  {k_4 J_3\cE_4J_3 k_4\over p_3k_4} \right\} \cM_3(123),
\ee
where $k_4 J_1\cE_4J_1 k_4 = J_1 \cR_4 J_1$ involves the linearised Riemann tensor, and thus it is manifestly gauge-invariant.
The $\pi^2$ factor form the expansion of the KLT kernel at $4$-point cancels exactly the $\zeta_2$ appearing in the expansion of the open string amplitudes,
thus implementing the single-valued projection discussed in Sec.~\ref{single-valued projection}.

\subsection{5-point amplitudes with massive states}   

Starting from the KLT expression for the $5$-point closed string amplitude
\begin{align}
 \cM_5(12345) &= \sin(\pi k_1p_2)  \sin(\pi p_3p_4) \cA^L_5(1[23]45) \cA^R_5(\underline{2}14\underline{3}5) \nonumber\\
&+\sin(\pi k_1p_3)  \sin(\pi p_2p_4) \cA_5^L(1[32]45) \cA_5^R(\underline{3}14\underline{2}5),
\end{align}
where we assume that $k_1^2=0$ (massless graviton) goes soft,  $k_1=\delta \hat{k}_1$, with $\delta{\rightarrow}0$. In this limit, we know that
\begin{align}
&\cA^L_5(12345) \approx \cS_1^{2-5} \cA^L_4(2345), \quad \quad 
\cA^R_5({2}14{3}5)  \approx \cS_1^{4-2} \cA^R_4({2}4{3}5), \nonumber\\
&\cA_5^L(13245)  \approx \cS_1^{3-5} \cA^L_4(3245), \quad \text{and}\quad
\cA_5^R({3}14{2}5) \approx \cS_1^{4-3} \cA^R_4({3}4{2}5).
\end{align}
Observing that 
\be
\sin(\pi p_3p_4) \cA^L_4(2345) \cA^R_4({2}4{3}5)  = \cM_4(2345) = \sin(\pi p_2p_4) \cA_4^L(3245) \cA_4^R({3}4{2}5),
\ee
one gets 
\be
 \cM_5(12345) \approx [\sin(\pi k_1p_2) \cS_1^{2-5} \cS_1^{4-2} + \sin(\pi k_1p_3) \cS_1^{3-5} \cS_1^{4-3}] \cM_4(2345).
\label{M5}
\ee
At leading order, Eq.~\eqref{M5} yields
 \be
 k_1p_2 \,\cS_{1(0)}^{2-5} \cS_{1(0)}^{4-2} + k_1p_3\, \cS_{1(0)}^{3-5} \cS_{1(0)}^{4-3}  = 
 \sum_{i\neq 1} {p_i \cE_1 p_i \over k_1 p_i} = \cS_{1(0)}^{grav}.
 \ee
 At sub-leading order
 \be
 k_1p_2 \left[\cS_{1(0)}^{2-5} \cS_{1(1)}^{4-2} + \cS_{1(1)}^{2-5} \cS_{1(0)}^{4-2}\right]
 + k_1p_3 \left[\cS_{1(0)}^{3-5} \cS_{1(1)}^{4-3} + \cS_{1(1)}^{3-5} \cS_{1(0)}^{4-3} \right]= 
 \sum_{i \neq 1} {k_1 J_i \cE_1 p_i \over k_1 k_i} = \cS_{1(1)}^{grav}.
 \ee
 At sub-sub-leading order 
 \begin{align}
& k_1p_2 \left[\cS_{1(0)}^{2-5}\cS_{1(2)}^{4-2} + \cS_{1(2)}^{2-5} \cS_{1(0)}^{4-2} + \cS_{1(1)}^{2-5} \cS_{1(1)}^{4-2}\right]
+  k_1p_3 \left[\cS_{1(0)}^{3-5} \cS_{1(2)}^{4-3} + \cS_{1(2)}^{3-5} \cS_{1(0)}^{4-3} + \cS_{1(1)}^{3-5} \cS_{1(1)}^{4-3} \right] \nonumber\\ 
 &\qquad\qquad\qquad \qquad \qquad=\sum_{i\neq 1} {k_1 J_i \cE_1 J_i k_1 \over k_1 p_i} = \sum_i {J_i \cR_1 J_i \over k_1 p_i} =\cS_{1(2)}^{grav},
 \end{align}
where the $\zeta_2$ factors coming from the KLT kernel cancel exactly those produced by the expansion at the sub-sub-leading of the 
$5$-point disk integral, as encoded by the single-valued projection.

\subsection{6-and higher-point amplitudes with massive states}   

Lastly, let us briefly focus on $6$-point amplitudes. In this case one has twelve terms
\begin{align}
\cM_6(123456) &=\sin(\pi k_1p_2)  \sin(\pi p_4p_5) \cA_L(1[234]56)  \nonumber\\
&\left\{ \sin(\pi k_1p_3) \cA_R(\underline{23}15\underline{4}6) + \sin(\pi p_3(k_1+p_2)) \cA_R(\underline{32}15\underline{4}6)\right\} + {\rm Perm} [234].
\end{align}
At leading order, we get 
\begin{align}
&\cM_6(123456) \approx \pi k_1p_2 \,\cS_1^{2-6}\cA_L(23456)  \nonumber\\
&\left[\sin(\pi p_4p_5) \sin(\pi p_3 p_2) 
\cS_{1(0)}^{5-2}\cA_R({32}5{4}6) + \sin(\pi p_3p_5) \sin(\pi p_4 p_2 )\cA_L(24356) \cS_{1(0)}^{5-2} \cA_R({42}5{3}6) \right] \nonumber\\
& \qquad \qquad \qquad \qquad \qquad \qquad \qquad \qquad \,\,+[2{\rightarrow}3] + [2{\rightarrow}4],
\end{align}
that yields
$$
\cM_6(123456) \approx \left\{ \pi \,k_1p_2 \, \cS_1^{2-6} \cS_1^{5-2} + [2{\rightarrow}3] + [2{\rightarrow}4]\right\} \cM_5(23456), 
$$
exposing the expected universal terms at leading order, where non-planarity is restored by summing over permutations in KLT or `single-valued map' formulae.
Sub-leading and sub-sub-leading are more laborious but are fixed by gauge invariance, as repeatedly discussed above. 

\section{Closed superstring amplitudes with massive insertions}
\label{superamps}

In this section we compute some amplitudes with insertions of massive string states.
Later on we will examine their soft behavior. 

Let us now consider closed superstrings and focus on the NS-NS sector. At the first massive level one finds a plethora of particles (all in all $2^{14} = 128\times 128 = (44 + 84) \times (44 + 84)$ d.o.f.) arising from the combinations $[H_{\mu\nu}\oplus C_{\mu\nu\rho}]_L\otimes[H_{\mu'\nu'}\oplus C_{\mu'\nu'\rho'}]_R$.  

\subsection{Three massless states one massive: $\cM_4(\cE_1,\cE_2,\cE_3,\cK_4+\cL_4+\cU_4)$}
Relying on KLT formulae one has 
\be
\cM(\cE_1,\cE_2,\cE_3,\cK_4+\cL_4+\cU_4) = \sin\left(\pi\frac{\alc}{4}t\right) \cA_L({A}_1,{A}_2,{A}_3,H_4+C_4)\cA_R({A}_1,{A}_3,{A}_2,H_4+C_4),
\ee
with $\cK+\cL +\cU = H\otimes \tilde{H}+C\otimes \tilde{C}+ H\otimes \tilde{C} + C\otimes \tilde{H}$.  The highest spin state is the Konishi top state with $s=4$ \cite{Bianchi:2001cm,Andrianopoli:1998ut,Eden:2005ve,Bianchi:2003wx}. In $D=10$ the explicit 
formula is extremely long and not very illuminating. We refrain for writing it down except for 
$\cL = C\otimes \tilde{C}$, whereby it reads 
\begin{align}
&\cM(\cE_1,\cE_2,\cE_3,\cK_4) ={\Gamma(1-\frac{s}{2}) \Gamma(1-\frac{t}{2}) \Gamma(1-\frac{u}{2}) \over \Gamma(\frac{t}{2}) \Gamma(1+\frac{u}{2}) \Gamma(1+\frac{s}{2})}  \nonumber \\
&\frac{us}{4} \,\biggr[C_4[a_1 a_2 a_3]+\sum_{i\neq 3}C_4[a_1 a_2 k_i]\frac{a_3k_i}{k_3k_i}+\sum_{i\neq 2}C_4[a_3 a_1 k_i]\frac{a_2k_i}{k_2k_i}+\sum_{i\neq 1}C_4[a_2 a_3 k_i]\frac{a_1k_i}{k_1k_i}\nonumber\\
&\qquad\qquad\qquad\qquad+C_4[a_1 k_2 k_3]\frac{a_2a_3}{k_2k_3}+C_4[a_2 k_3 k_1]\frac{a_3a_1}{k_3k_1}+C_4[a_3 k_1 k_2]\frac{a_1a_2}{k_1k_2}\biggr]^2.
\label{MEEEL}
\end{align}
We shall also study the soft behavior of the amplitude $\cM(\cE_1,\cE_2,\cE_3,\cK_4)$.
It is worth to notice that $\cK_4=H\otimes \widetilde{H}$ is a reducible tensor. The following decomposition holds
\begin{align}
{\bf 44}\otimes {\bf 44}&={\bf 450} \oplus {\bf 910} \oplus {\bf 495} \oplus {\bf 44} \oplus {\bf 36} \oplus {\bf 1},\\
(2,0)\otimes(2,0)&=(4,0)\oplus(2,1)\oplus(0,2)\oplus(2,0)\oplus(0,1)\oplus(0,0).
\end{align}
In particular, this product contains the $10$-dimensional analogue of a spin 4 state 
\be
S_{\mu_1\mu_2\mu_3\mu_4}=1/2(H_{\mu_1\mu_2}\widetilde{H}_{\mu_3\mu_4}+H_{\mu_3\mu_4}\widetilde{H}_{\mu_1\mu_2})-1/(9\times 4)\sum_{i,k=1,2}\sum_{j,l=3,4}H_{\lambda\mu_i}\widetilde{H}^\lambda_{\mu_j}\delta_{\mu_k\mu_l}.
\ee

In $D=4$ the situation drastically simplifies. Focussing on the combinations of the $SO(6)$ singlets $H_{\mu\nu} = H^{tt}_{\mu\nu} + H_0(\eta_{\mu\nu} + \ap p_\mu p_\nu)$ (with $H_{ij} ={-}H_0\delta_{ij}/2$) and $C_{\mu\nu\rho} = C_0 \sqrt{\ap} p^\lambda\varepsilon_{\lambda\mu\nu\rho}$ that couple to two gluons, one has 49 d.o.f. that assemble in five scalars, one vector, five spin-2 (5 states each), one spin 3 (7 states) and one spin 4 (9 states).  
Since the $H^{tt}_{\mu\nu}$ couples to gluons with opposite helicity while $H_0/C_0$ couple to gluons with the same helicity, the open-string building blocks are 
$$
\cA(1^-,2^+,3^+,H^{++}) \quad , \quad  \cA(1^+,2^+,3^+,H_{0}/C_0) \quad , \quad \cA(1^-,2^-,3^+,H_{0}/C_0)
$$
and the ones related to them by Lorentz transformations (acting on $H^h$), conjugation or permutations of the gluons. 

For instance, the amplitude of 3 gravitons with the top component $K^{+4} = u_4^4 \bar{v}_5^4$ (recall $p_4 = k_4 + k_5 = u_4\bar{u}_4 + v_5\bar{v}_5$) reads
\begin{align}
&\cM_4(1^{-2},2^{+2},3^{+2},K_4^{+4})=\sin\left(-\pi k_2k_3\right)\times \cA_L(1^-2^+3^+H_4^{+2})\otimes \cA_R(1^-3^+2^+H_4^{+2})\nonumber\\
&=G_N\pi\frac{\Gamma\left( k_3p_4\right)\Gamma\left(1+ k_2k_3\right)\Gamma\left(1+ k_1 k_3 \right)}{\Gamma\left(2- k_3p_4\right)\Gamma\left(- k_2k_3\right)\Gamma\left(1- k_1k_3\right)}\frac{[13]\lef 14 \re^4 [45]^2}{\lef 12 \re \lef 23 \re m_H^3}\frac{[12]\lef 14 \re^4 [45]^2}{\lef 13 \re \lef 32 \re m_H^3}.
\label{MGGGK}
\end{align}
The amplitudes for the lower spin components of $\cK$ follow performing $SO(3)$ little group transformations on the above one. Similarly one can replace two gravitons with dilatons or axions
\begin{align} &\cM_4(1^{0}2^{0}3^{+2}K^{+4})  =  \sin(-\pi k_2k_3) \cA_L(1^{-}2^{+}3^{+}H^{++}) \otimes \cA_R(1^{+}3^{+}2^{-}H^{++}) \nonumber \\
&=
G_N\pi\frac{\Gamma\left( k_3p_4\right)\Gamma\left(1+ k_2k_3\right)\Gamma\left(1+ k_1 k_3 \right)}{\Gamma\left(2- k_3p_4\right)\Gamma\left(- k_2k_3\right)\Gamma\left(1- k_1k_3\right)} \frac{ [13] \lef 14 \re^4 [45]^2}{\lef 12 \re \lef 23 \re m_H^3}  \frac{ [12] \lef 24 \re^4 [45]^2}{ \lef 13 \re \lef 32 \re m_H^3}
\label{MDDGK}
\end{align}
Once again, amplitudes for the other helicity states of $\cK$ obtain after $SO(3)$ little group transformations. Note, for instance, that $\cK_{s=4}^{+++0} = H^{++}\tilde{H}^{+0}+ H^{+0}\tilde{H}^{++}$ while $\cK_{s=3}^{+++} = H^{++}\tilde{H}^{+0}- H^{+0}\tilde{H}^{++}$. The former is even under $\Omega$, the latter is odd.

One can also consider the 4 real (2 complex) $s=2$ massive states corresponding to $H^0/C^0 \otimes \tilde{H}^{2} \pm  H^2 \otimes \tilde{H}^{0}/\tilde{C}^0$ whose amplitudes with massless states obtain from combinations of  $\cA(1^-,2^+,3^+,H^{++})$ with $\cA(1^+,2^+,3^+,H_{0}/C_0)$ or $\cA(1^-,2^-,3^+,H_{0}/C_0)$. For instance
\begin{align} 
\cM(1^{0}2^{+2}3^{+2}\cH^{+2})  &=  \sin\left(-\pi k_2k_3\right) \cA_L(1^{-}2^{+}3^{+}H^{++}) \cA_R(1^{+}3^{+}2^{+}H^{0}/C^0) \nonumber \\
&=G_N\pi\frac{\Gamma\left( k_3p_4\right)\Gamma\left(1+ k_2k_3\right)\Gamma\left(1+ k_1 k_3 \right)}{\Gamma\left(2- k_3p_4\right)\Gamma\left(- k_2k_3\right)\Gamma\left(1- k_1k_3\right)} \frac{ [13] \lef 14 \re^4[45]^2}{ \lef 12 \re \lef 23 \re m_H^3}  
\frac{ [12]m_H^3 }{\lef 13 \re \lef 32 \re}
\label{MDGGK}
\end{align}
Finally amplitudes for the four scalars $H^0/C^0 \otimes \tilde{H}^{0}/\tilde{C}^0$  obtain combining 
$\cA(1^+,2^+,3^+,H_{0}/C_0)$ or 

$\cA(1^-,2^-,3^+,H_{0}/C_0)$ with each other and with permutations thereof. 

\section{Bosonic string amplitudes with `massive' insertions}
\label{boseamps}
\subsection{Three-tachyons one-massless: $\cM_4(\Tac_1,\Tac_2,\cE_3,\Tac_4)$}

Consider now also the insertion of generic massless closed string states with $k^2=0$ 
\be
V_\cE(z,\bar z)=\cE_{\mu\nu} i\de X_L^\mu(z) i\bar\de X_R^\nu(\bar z) e^{i{k\over 2}X_L(z)} e^{i{k\over 2}X_R(\bar z)}, 
\ee
where $\cE_{\mu\nu}$ is transverse with respect to both indices $k^\mu \cE_{\mu\nu} = 0 = k^\nu \cE_{\mu\nu}$. Decomposing $\cE_{\mu\nu} = h_{\mu\nu} + \phi_{\mu\nu} + b_{\mu\nu}$ into irreducible representations of the Lorentz group, 
$ h_{\mu\nu} = h_{\nu\mu} $ with $\eta^{\mu\nu}h_{\nu\mu} = 0$ describes the graviton,
$\phi_{\mu\nu} = \eta_{\mu\nu} -k_{\mu}\bar{k}_{\nu} -k_{\nu}\bar{k}_{\mu}$ with $\bar{k}\bar{k} = 0$ and  ${k}\bar{k} = 1$ describes the dilaton and $b_{\mu\nu}= - b_{\nu\mu}$ the Kalb-Ramond field.
Consider the amplitude:
\begin{align}
&\cM_4(\Tac_1,\Tac_2,\cE_3,\Tac_4)\nonumber\\
&=\left\langle c\bar{c}\,e^{ip_1X}(z_1,\bar z_1)\,\,c\bar{c}\,e^{ip_2X}(z_2,\bar z_2)\,\,\int \frac{d^2z_3}{\pi}i\de X\,\cE\,i\bar\de X\,e^{ik_3X}(z_3,\bar z_3)\,\,c\bar{c}\,e^{ip_4X}(z_4,\bar z_4)\right\rangle\nonumber\\
&=\int \frac{d^2z}{\pi}\,P_3\cE_3 \bar P_3 |z|^{2 p_3p_4}|1-z|^{2 p_2p_3}.
\end{align}
Since
\be
P_3=\frac{p_1}{z_{31}}+\frac{p_2}{z_{32}}+\frac{p_4}{z_{34}} \overset{z_1\to\infty}{\rightarrow}\frac{p_4}{z}-\frac{p_2}{1-z},
\ee
the amplitude reads as
\begin{align}
&\int \frac{d^2z}{\pi}\, |z|^{2 p_3p_4}|1{-}z|^{2 p_2p_3} \left( \frac{p_4\cE_3p_4}{|z|^2} {+} \frac{p_2\cE_3p_2}{|1{-}z|^2}{-}\frac{p_2\cE_3p_4}{\bar z(1{-}z)}{-}\frac{p_4\cE_3p_2}{z(1{-}\bar z)} \right)\nonumber\\
&=p_4\cE_3p_4\,\cI(2 p_3p_4 {-}2,0;2 p_2p_4,0){+}p_2\cE_3p_2\,\cI(2 p_3p_4,0;2 p_2p_3{-}2,0)\nonumber\\
&{-}(p_2\cE_3p_4{+}p_4\cE_3p_2)\,\cI(2 p_3p_4{-}2,1;2 p_2p_3,{-}1)\nonumber\\
&=\frac{1}{k_3p_1\,\,k_3p_2\,\,k_3p_4}({-}p_4\cE_3p_4 (k_3p_2)^2 {-} p_2\cE_3p_2 (k_3p_4)^2 {+} p_2(\cE_3{+}\cE_3^t)p_4\,\,k_3p_4\,\,k_3p_2)\nonumber\\
&\frac{\Gamma(1{+} k_3p_4)\Gamma(1{+} k_3p_2)\Gamma(1{+} k_3p_1)}{\Gamma(1{-} k_3p_4)\Gamma(1{-} k_3p_2)\Gamma(1{-} k_3p_1)}\nonumber\\
&=\frac{1}{k_3p_1\,\,k_3p_2\,\,k_3p_4}(p_1\cE_3p_4 (k_3p_2)^2 {+} p_1\cE_3p_2 (k_3p_4)^2{+}p_2\cE_3p_4 (k_3p_1)^2)\nonumber\\
&\frac{\Gamma(1{+} k_3p_4)\Gamma(1{+} k_3p_2)\Gamma(1{+} k_3p_1)}{\Gamma(1{-} k_3p_4)\Gamma(1{-} k_3p_2)\Gamma(1{-} k_3p_1)}.
\end{align}

One concludes that only the symmetric part of $\cE_S = \frac{1}{2}({\cal E}_3+{\cal E}_3^t)$ contributes due to symmetry under world-sheet parity $\Omega$, under which $h$ and $\phi$ are even while $b$ is odd.

The $\Gamma$ functions in the above expression can be rearranged as
\be
\widetilde{B}(1,2,\hat 3,4)\widetilde{B}(4,2,\hat 3,1)\sin\left(\pi k_3p_2\right),
\ee
where
\be
\widetilde{B}(1,2,\hat 3,4)=\frac{\Gamma(1+ k_3p_2)\Gamma(1+ k_3p_4)}{\Gamma(1- k_3p_1)}.
\ee
Moreover
\be 
-p_4\cE_3p_4 (k_3p_2)^2 - p_2\cE_3p_2 (k_3p_4)^2 + p_2(\cE_3+\cE_3^t)p_4\,\,k_3p_4\,\,k_3p_2
\nonumber \ee
\be
= -(p_4a_3\, k_3p_2 - p_2a_3 \, k_3p_4) (p_1\tilde{a}_3 \, k_3p_2 - p_2\tilde{a}_3\, k_3p_1) \ee
so much so that
\be
\cM_4(\Tac_1,\Tac_2,\cE_3,\Tac_4) = \sin\left(\pi\frac{t}{2}\right) \cA^L_4(T_1,T_2,A_3,T_1)\cA^R_4(T_4,T_2,A_3,T_1),
\ee
as expected.

\subsection{Two-tachyons two-massless: $\cM_4(\cE_1,\cE_2,\Tac_3,\cT_4)$}

Using KLT in the s-channel (1-2 or 3-4 exchange) one finds
\be
\cM(\cE_1,\cE_2,\Tac_3,\cT_4) =  \sin\left(\pi\frac{s}{2}\right)\cA^L_4({A}_1,{A}_2, T_3,T_4)
\cA^R_4({A}_1, {A}_2, T_4, T_3) 
\ee
so that the two-massless two-tachyon amplitude reads
\begin{align}
&\cM(\cE_1,\cE_2,\Tac_3,\cT_4) = \frac{\Gamma(1+ k_1p_3)\Gamma(1+ k_1p_4)\Gamma(-1+ k_1k_2)}{\Gamma(- k_1p_3)\Gamma(- k_1p_4) \Gamma(2- k_1k_2)} \\
&\biggr(a_1a_2-(a_1p_3\,\,a_2p_3 + a_1p_4\,\,a_2p_4) +a_1p_3\,\,a_2p_4\frac{1+ k_1p_4}{k_1p_3}+a_1p_4\,\,a_2p_3\frac{1+ k_1p_3}{k_1p_4}\biggr) \nonumber \\
&\biggr(\tilde{a}_1\tilde{a}_2-(\tilde{a}_1p_3\,\,\tilde{a}_2p_3 + \tilde{a}_1p_4\,\,\tilde{a}_2p_4) +\tilde{a}_1p_3\,\,\tilde{a}_2p_4\frac{1+ k_1p_4}{k_1p_3} +\tilde{a}_1p_4\,\,\tilde{a}_2p_3\frac{1+ k_1p_3}{k_1p_4}\biggr)
\label{M4TTET}
\end{align}
Replacing $a_i^\mu \tilde{a}^\nu_i = \cE^{\mu\nu}_i$ one gets 
$$
\cM(\cE_1,\cE_2,\Tac_3,\cT_4) = \cI(s,t,u)\cE^{\mu\nu}_1 \cE^{\rho\sigma}_2 K_{\mu\rho} K_{\nu\sigma}
$$
where
\be
 \cI(s,t,u) = \frac{\Gamma(1+ k_1p_3)\Gamma(1+ k_1p_4)\Gamma(-1+ k_1k_2)}{\Gamma(- k_1p_3)\Gamma(- k_1p_4) \Gamma(2- k_1k_2)} 
\label{dynMTTET}
 \ee
and 
$$
K^{\mu\nu} = \eta^{\mu\nu} - (p^\mu_3 p^\nu_3 + p^\mu_4p^\nu_4) +p^\mu_3p^\nu_4\frac{1+ k_1p_4}{k_1p_3} +p^\mu_4p^\nu_3\frac{1+ k_1p_3}{k_1p_4}
$$
that shows that only $\cM(h/\phi_1, h/\phi_2, T_3,T_4)$ and $\cM(b_1, b_2, T_3,T_4)$ are non-vanishing, as expected on the basis of world-sheet parity symmetry $\Omega$.

\subsection{Two-tachyons one-massless one-massive: $\cM_4(\Tac_1,\Tac_2,\cE_3,\cK_4)$}

Using KLT in the s-channel (1-2 exchange) one finds
\be
\cM(\Tac_1,\cT_2,\cE_3,\cK_4) =  \sin\left(\pi\frac{s}{2}\right)\cA_L(T_1,T_2,{A}_3,H_4)\otimes\cA_R(T_2,T_1,{A}_3,H_4) 
\ee
or more explicitly
\begin{align}
&\cM(\Tac_1,\Tac_2,\cE_3,\cK_4) = \frac{\Gamma(1{+} p_1k_3)\Gamma({-}1{+}k_3p_4)\Gamma(1{+} p_2k_3)}{\Gamma({-} p_1k_3) \Gamma(2{-} k_3p_4) \Gamma({-} p_2k_3)} \nonumber\\
&\biggr[{-}2a_3Hp_2{-}2a_3Hk_3\frac{1{+} k_3p_1}{2{-} k_3p_4} 
{+} a_3p_4\biggr(p_2Hp_2\frac{1{-} k_3p_4}{ k_3p_1} {+}k_3Hk_3\frac{1{+}k_3p_1}{2{-}k_3p_4}{+}2p_2Hk_3\biggr) \nonumber\\
&{-} a_3p_2\biggr(\frac{k_3p_4\,\,p_2Hp_2}{ p_2k_3\,\,p_1k_3}(1{-} k_3p_4)
{-}k_3Hk_3\frac{1{+} p_1k_3}{ p_2k_3}{-}2p_2Hk_3\frac{1{-} k_3p_4}{ p_2k_3}\biggr)\biggr] \nonumber \\
&\otimes\biggr[{-}2\tilde{a}_3\tilde{H}p_1{-}2\tilde{a}_3\tilde{H}k_3\frac{1{+} k_3p_2}{2{-} k_3p_4} 
{+}\tilde{a}_3p_4\biggr(p_1\tilde{H}p_1\frac{1{-} k_3p_4}{ k_3p_2} {+}k_3\tilde{H}k_3\frac{1{+} k_3p_2}{2{-} k_3p_4}{+}2p_1\tilde{H}k_3\biggr) \nonumber\\
&{-} \tilde{a}_3p_1\biggr(\frac{k_3p_4\,\,p_1\tilde{H}p_1}{ p_1k_3\,\,p_2k_3}(1{-} k_3p_4)
{-}k_3\tilde{H}k_3\frac{1{+} p_2k_3}{ p_1k_3}{-}2p_1\tilde{H}k_3\frac{1{-} k_3p_4}{ p_1k_3}\biggr)\biggr],
\label{MTTEH}
\end{align}
where $\cE_3=a_3\otimes\tilde{a}_3$ and  $\cK_4=H\otimes\tilde{H}$. Without much effort one can check that $\cE_{\pm}=a\otimes\tilde{a}\pm\tilde{a}\otimes{a}$ with definite parity under $\Omega$ couple to $\cK_{\pm}=H\otimes\tilde{H}\pm \tilde{H}\otimes{H}$ with the same parity.

\section{Soft limit of closed string amplitudes with massive insertions}
\label{softexamples}
In this section, we study the soft limit of 4-point amplitudes with massive insertions. We start with the superstring and focus on the $D=4$ case where the spinor helicity formalism largely simplifies the results. We then pass to consider the bosonic strings and study tachyon insertions, too. Finally we investigate the soft behaviour for amplitudes with two Kalb-Ramond fields. 

\subsection{Soft limit of superstring amplitudes in the spinor helicity formalism}

Restring the momenta and polarisations to $D=4$ allows us to derive compact expressions for the universal soft operator in the spinor helicity formalism. For simplicity we focus on 4-point amplitudes with three massless and one massive external legs. 
In particular, we will consider the soft limit of the amplitudes in Eqs.~\eqref{MGGGK},~\eqref{MDDGK}, and~\eqref{MDDGK}, computed using KLT. 
When the graviton with helicity $h=+2$ and momentum $k_3$ goes to zero, we find
\begin{align}
\cS^{0}&{=}\sqrt{G_N}\frac{[13][23]\lef 12 \re^2}{\lef 13 \re \lef 32 \re 2k_3p_4}\\
\cS^{1}&{=}\sqrt{G_N}\frac{1}{\lef 3q_3 \re}\biggr[\frac{\lef 1q_3 \re [31]}{\lef 13 \re}\widetilde u_3\frac{\partial}{\de \widetilde u_1} {+}\frac{\lef 2q_3 \re [32]}{\lef 23 \re}\widetilde u_3\frac{\partial}{\de \widetilde u_2} {+}\frac{\lef 4q_3 \re [43]{+}\lef 5q_3 \re [53]}{2k_3p_4}\nonumber\\
&\quad\quad\quad\quad\,\left( [34]\widetilde u_3\frac{\de}{\de \widetilde u_4}{+}  [35]\widetilde u_3\frac{\de}{\de \widetilde u_5} \right) \biggr]\\
\cS^{2}&{=}\sqrt{G_N}\left[\frac{[13]}{2\lef31\re}\left(\widetilde u_3\frac{\de}{\de \widetilde u_1}\right)^2{+}\frac{[23]}{2\lef32\re}\left(\widetilde u_3\frac{\de}{\de \widetilde u_2}\right)^2{+}\frac{1}{4k_3p_4}\left([34]\widetilde u_3 \frac{\de}{\de\widetilde u_4}{+}[35]\widetilde u_3 \frac{\de}{\de\widetilde u_5}\right)^2\right].
\label{holosoft}
\end{align}
Applying the operators $\cS^{i},\,i=0,1,2$ to the amplitudes
\begin{align}
\cM_3(1^{-2},2^{+2},\cK_4^{+4})&=\sqrt{G_N}\frac{\lef 14 \re^4[25]^4}{m^6}\\
\cM_3(\phi_1,\phi_2,\cK_4^{+4})&=\sqrt{G_N}\frac{\lef 14 \re^2 \lef 24 \re^2 [15]^2 [25]^2}{m^6}\\
\cM_3(\phi_1,2^{+2},\cH_4^{+4})&=\sqrt{G_N}\frac{\lef 14 \re^2 \lef 24\re^2 [12]^2}{m^4},
\end{align}
we reproduce the soft expansions found respectively in~\ref{appendixGGGK},~\ref{appendixDDGK}, and~\ref{appendixDGGK}
\begin{align}
\cS^0\cM_3(1^{-2}2^{+2}K_4^{+4})&=G_N\frac{[13][23]\lef 12 \re^2}{\lef 13 \re \lef 32 \re 2k_3p_4}\frac{\lef 14 \re^4[25]^4}{m^6}\\
\cS^1\cM_3(1^{-2}2^{+2}K_4^{+4})&=4G_N\frac{[13][23][35][25]^3\lef 12 \re \lef 14 \re^4}{\lef 32 \re m^6 2k_3p_4}\\
\cS^2\cM_3(1^{-2}2^{+2}K_4^{+4})&=6G_N\frac{[13][23][35]^2[25]^2\lef 13 \re\lef 14 \re^4}{\lef 32 \re m^6 2k_3p_4}.
\end{align}
Had we chosen the leg with momentum $k_1$ to be soft in Eq.~\eqref{MGGGK}, we would have gotten a trivial result, since the interaction vertex vanishes $\cM_3(\cE_2^{+2},\cE_3^{+2},\cK_4^{+4})=0$. While our results are symmetric in the exchange of $2\leftrightarrow 3$, when the external leg with momentum $k_2$ is a graviton.

\subsection{Soft limit of bosonic string amplitudes}
\subsubsection{$\cM_4(\Tac_1,\Tac_2,\cE_3,\Tac_4)$}
The simplest case to be considered is the amplitude with three tachyons and the one graviton $\cM_4(\Tac_1,\Tac_2,\cE_3,\Tac_4)$
\begin{align}
&\cM_4(\Tac_1,\Tac_2,\cE_3,\Tac_4)=\frac{\Gamma(1{{+}}k_3p_4)\Gamma(1{{+}} k_3p_2)\Gamma(1{{+}} k_3p_1)}{\Gamma(1{{-}} k_3p_4)\Gamma(1{{-}} k_3p_2)\Gamma(1{-}k_3p_1)}
\left( \frac{p_1\cE_3 p_1}{k_3p_1} {{+}} \frac{p_2\cE_3p_2}{k_3p_2} {+} \frac{p_4\cE_3p_4}{k_3p_4} \right).
\end{align}
The dynamical factor in the above expression has a very special soft behavior
\begin{align}
&\frac{\Gamma(1{+} k_3p_4)\Gamma(1{+}k_3p_2)\Gamma(1{+}k_3p_1)}{\Gamma(1{{-}}k_3p_4)\Gamma(1{-}k_3p_2)\Gamma(1{-} k_3p_1)}\nonumber\\
&=\frac{1{+}\sum_{i\neq 3}k_3p_i\Gamma^\prime(1){+}\frac{1}{2}\sum_{i\neq 3}(k_3p_i)^2\Gamma^{\prime\prime}(1){+}\sum_{i<j;i,j\neq 3}k_3p_i\,k_3p_j\Gamma^{\prime 2}(1){+}\cO(\delta^3)}{1{-}\sum_{i\neq 3}k_3p_i\Gamma^\prime(1){+}\frac{1}{2}\sum_{i\neq 3}(k_3p_i)^2\Gamma^{\prime\prime}(1){+}\sum_{i<j;i,j\neq 3}k_3p_i\,k_3p_j\Gamma^{\prime 2}(1){+}\cO(\delta^3)}\nonumber\\
&\qquad \qquad \qquad \qquad \qquad\qquad\qquad\qquad\qquad\qquad\qquad\qquad\qquad\qquad\quad=1{+}\cO(\delta^3).
\label{Gamma3}
\end{align}
Eq.~\eqref{Gamma3} does not spoil the soft behavior of the amplitude up to the sub-sub-leading order.
This happens every time the dynamical factor depends on the soft momentum as in Eq.~\eqref{Gamma3} and in all cases we are going to study we will 
always extract this factor.
At this stage, the expansion of the amplitude yields
\be
\cM_4(\Tac_1,\Tac_2,\cE_3,\Tac_4)= \frac{p_1\cE_3 p_1}{k_3p_1} {+} \frac{p_2\cE_3p_2}{k_3p_2} + \frac{p_4\cE_3p_4}{k_3p_4} +\cO(\delta^3).
\ee
Which agrees with the expected soft behavior since the three amplitude $\cM_3(\Tac_1,\Tac_2,\Tac_4)$ is just a number, 
so the action of the angular momentum operator gives zero. 

\subsubsection{$\cM_4(\cE_1,\cE_2,\Tac_3,\Tac_4)$}

When $\cE_1=h_1/\phi_1$ and $\cE_2=h_2/\phi_2$ the soft theorem would suggest the following expansion for the amplitude in Eq.~\eqref{M4TTET}
\begin{align}
\cS^{0}\cM_3(\cE_2,\Tac_3,\Tac_4)&=\left( \frac{p_3\cE_1p_3}{k_1p_3} +\frac{p_4\cE_1p_4}{k_1p_4}+\frac{k_2\cE_1k_2}{k_1k_2}\right)\frac{p_-}{2}\cE_2\frac{p_-}{2};\\
\cS^{1}\cM_3(\cE_2,\Tac_3,\Tac_4)&{=}\left(\frac{k_1J_2\cE_1k_2}{k_1k_2}+\frac{k_1J_3\cE_1p_3}{k_1p_3}+\frac{k_1J_4\cE_1p_4}{k_1p_4}\right)\frac{p_-}{2}\cE_2\frac{p_-}{2}\nonumber\\
&{=}\frac{k_2\cE_1p_-\,k_1\cE_2p_- {-}k_1p_-\,k_2\cE_1\cE_2p_-}{2k_1k_2}+\frac{p_3\cE_1\cE_2p_{-}\,k_1p_3{-}p_3\cE_1p_3\,k_1\cE_2p_-}{2k_1p_3} \nonumber\\
&{+}\frac{p_4\cE_1p_4\,k_1\cE_2p_-{-}p_4\cE_1\cE_2p_-\,k_1p_4 }{2k_1p_4};\\
\cS^{2}\cM_3(\cE_2,\Tac_3,\Tac_4)&=\left(\frac{k_1J_2\cE_1J_2k_1}{2k_1k_2}+\frac{k_1J_3\cE_1J_3k_1}{2k_1p_3}+\frac{k_1J_4\cE_1J_4k_1}{2k_1p_4}\right)\frac{p_-}{2}\cE_2\frac{p_-}{2}\nonumber\\
&{=}\frac{ k_1p_-\,p_-\cE_1\cE_2k_1{-}p_-\cE_1p_- ~k_1\cE_2k_1{-}\cE_1\cE_2 (k_1p_-)^2}{4k_1k_2}\nonumber\\
&{+}\frac{2p_3\cE_1\cE_2k_1\,k_1p_3{-}k_1\cE_2k_1\,p_3\cE_2p_3{-}(k_1p_3)^2\,\cE_1\cE_2}{4k_1p_3}\nonumber\\
&{+}\frac{2p_4\cE_1\cE_2k_1\,k_1p_4{-}k_1\cE_2k_1\,p_4\cE_2p_4{-}(k_1p_4)^2\,\cE_1\cE_2}{4k_1p_4}.
\end{align}
Since Kalb-Ramond $b$-fields are odd under world-sheet parity we would expect zero because $\cM_3(b_2,\Tac_3,\Tac_4)=0$.
Following the steps reported in~\ref{appendixTTET} we find that at the sub-sub-leading order the soft behavior of the amplitude is not 
reproduced by the soft operator $\cS^2$. In particular, there are additional terms that we expect coming from the $\cM_3(h_1,h_2,\phi_I)$ vertex, Eq.~\eqref{TTETadditional}.
    
For two Kalb-Ramond fields $\cE_{1,2}=b_{1,2}$ the amplitude at leading order $\cO(\delta^{-1})$ is zero. 
The expansion starts at order $\cO(\delta^0)$ 
\begin{align}
\cM^{(0)}_4(b_1,b_2,\Tac_1,\Tac_2)&=\frac{1}{2}p_-b_1b_2p_-{+}\frac{k_1b_2p_-\,k_2b_1p_-}{2k_1k_2}+\frac{k_1p_-\,k_2b_1b_2p_-}{2k_1k_2}\\
\cM^{(1)}_4(b_1,b_2,\Tac_1,\Tac_2)&={-}\frac{1}{2}k_1p_-\,k_2b_1b_2p_-{+}\frac{1}{2}k_1b_2p_-\,k_2b_1p_-{-}\frac{1}{2}k_2b_1b_2k_1\nonumber\\
&+\frac{k_1p_-\,p_-b_1b_2k_1}{2k_1k_2}+\frac{1}{2}k_1k_2\,p_-b_1b_2p_-+\frac{1}{4}k_1k_2\text{tr}(b_1b_2)-\frac{(k_1p_-)^2\text{tr}(b_1b_2)}{4k_1k_2}
\end{align}

It is worth to notice that there are only poles in $k_1k_2$, as expected since $\cM_3(b,\cT,\cT)=0$ due to world-sheet parity.
One can try to interpret the soft result as a factorization on the massless pole {\it viz.}
\be
\lim_{k_1\rightarrow 0} \cM_4(b_1,b_2,\cT_3,\cT_4) = \sum_{e(k)}\cM_3(b_1,b_2,e(-k_1-k_2)) {1\over 2 k_1k_2} \cM_3(e(k_1+k_2),\cT_3,\cT_4)
\ee
where $e(k)$ collectively denotes the physical polarisations of the graviton and dilaton $e_{\mu\nu} = h_{\mu\nu} + \phi_{\mu\nu}$. Alternatively, since $2 k_1k_2 = - 2 (k_1+k_2)p_3 = - 2 (k_1+k_2)p_4$, one can envisage 
a `double soft limit', see \emph{e.g.}~\cite{ArkaniHamed:2008gz,Chen:2014xoa,Chen:2014cuc},
\be
\lim_{k_1, k_2\rightarrow 0} \cM_{n+2}(b_1,b_2,H_3,..., H_{n+2}) = \sum_i {1\over (k_1+k_2)p_{i}} \cD(b_1,b_2; k_1-k_2)  \cM_{n}(H_3,..., H_{n+2})
\ee
where our present computations suggest
\be
\cD(b_1,b_2; k_1-k_2) = (k_2-k_1)b_1P_i \, (k_1-k_2)b_2P_i + {1 \over 4} [(k_1-k_2)^2 P_i - (k_1-k_2)P_i 
(k_1-k_2)] \{b_1,b_2\}P_i
\ee
Clearly this issue deserves further investigation.~\footnote{We thank Paolo Di Vecchia and Raffaele Marotta for interesting and fruitful discussions on this and related issues.}

\subsubsection{$\cM_4(\Tac_1,\Tac_2,\cE_3,\cK_4)$}

For simplicity we consider only the case in which $\cK_4[\mu,\nu,\rho,\sigma]$ is the completely symmetric irreducible state. In this case due to $\Omega$-parity $\cE_3=h_3/\phi_3$ only.
Applying the soft operators to the three level amplitude 
\be
\cM_3(\Tac_1,\Tac_2,\cK_4)=\cK_4\left[\pmm,\pmm,\pmm,\pmm\right]
\ee
we expect the following behavior 
\begin{align}
\cS^{0}\cM_3(\Tac_1,\Tac_2,\cK_4)&=\left(\frac{p_1\cE_3p_1}{k_3p_1}+\frac{p_2\cE_3p_2}{k_3p_2}+\frac{p_4\cE_3p_4}{k_3p_4}\right)\cK_4\left[ \pmm,\pmm,\pmm,\pmm \right];\\
\cS^{1}\cM_3(\Tac_1,\Tac_2,\cK_4)&=\left(\frac{p_1\cE_3J_1k_3}{k_3p_1}{+}\frac{p_2\cE_3J_2k_3}{k_3p_2}{+}\frac{p_4\cE_3J_4k_3}{k_3p_4}\right)\cK_4\left[ \pmm,\pmm,\pmm,\pmm \right]\nonumber\\
&=2\left(\frac{p_1\cE_3p_1}{k_3p_1}-\frac{p_2\cE_3p_2}{k_3p_2}-\frac{p_-\cE_3p_4}{k_3p_4}\right)\cK_4\left[\pmm,\pmm,\pmm,k_3\right]\nonumber\\
&+2\left(-(p_1\cE_3)^\mu+(p_2\cE_3)^\mu+\frac{(p_4\cE_3)^\mu}{k_3p_4}k_3p_-\right)\cK_4\left[ _\mu,\pmm,\pmm,\pmm \right];\\
\cS^{2}\cM_3(\Tac_1,\Tac_2,\cK_4)&=\frac{3}{2}\left((p_1\cE_3)^\mu+(p_2\cE_3)^\mu +\frac{k_3p_-\,(p_-\cE_3)^\mu}{k_3p_4} \right)\cK_4\left[_\mu,k_3,\pmm,\pmm\right]\nonumber\\
&-\frac{3}{4}\left(k_3p_1\,+k_3p_2 +\frac{(k_3p_-)^2}{k_3p_4}\right)\cK_4\left[\cE_3,\pmm,\pmm\right]\nonumber\\
&-\frac{3}{4}\left( \frac{p_1\cE_3p_1}{k_3p_1}+\frac{p_2\cE_3p_2}{k_3p_2}+\frac{p_-\cE_3p_-}{k_3p_4}\right)\cK_4\left[k_3,k_3,\pmm,\pmm\right],
\end{align}
where
\be
J_{4\mu\nu}=p_{4[\mu}\frac{\de}{\de p_4^{\nu ]}}+4\cK_4[_\mu,_\alpha,_\beta,_\gamma]\frac{\de}{\de \cK_4[^\nu,_\alpha,_\beta,_\gamma]}.
\ee
Following the steps outlined in~\ref{appendixTTEK}, we reproduce the leading and sub-leading behavior as predicted by the soft theorem, but not 
the sub-sub-leading order. As for the amplitude $\cM_4(\cE_1,\cE_2,\Tac_3,\Tac_4)$ we are led to think that the mixing with the other degenerate string states 
spoil the soft theorem statement at this order.

\section{Conclusions and outlook}
\label{conclusions}

We have extended our analysis of the soft behaviour of string amplitudes with massive insertions to closed strings. Relying on our previous results for open strings and on KLT formulae we have checked universality of the soft behaviour to sub-leading order for superstring amplitudes. At sub-sub-leading order we have argued in favour of universality on the basis of OPE of massless and massive vertex operators and gauge invariance with respect to the soft gravitons. We have also checked our statements against explicit 4-point amplitudes with one massive insertion in any dimension, including $D=4$, where use of the helicity spinor formalism drastically simplifies all expressions. As a by-product of our analysis we have checked the cancellation of $\pi^2$ arising from $\sin(\pi\alpha^\prime_{c} k_ik_j)$ factors in KLT formula with those arising from open superstring amplitudes in the soft limit, at sub-sub-leading order. This is expected for the `single valued projection' advocated in~\cite{StStSVP,StStTaylor} to hold for massive amplitudes, too.
This is comforting, being closed string theory of quantum gravity. Yet, our results are only valid at tree level and the proper extension to one- and higher-loops is still under debate in that IR divergences seem to produce non-universal $\log\delta$ terms~\cite{TravaNew} even in ${\cal N} =4$ SYM at one-loop, let alone supergravity or superstring theories. It would be very interesting to investigate this subject along the lines of~\cite{MBAVS, MBDario} and establish whether $\log\delta$ terms exponentiate, as usual for IR divergences, and in case which would be the relevant `anomalous' dimension that governs this hopefully universal behaviour. The approach proposed in~\cite{Avery:2015gxa,AS} based on the second N\"other theorem seems promising in this respect, though so far shown to be valid only at tree level.
 
\medskip
\noindent\textbf{Acknowledgments}
\medskip

We would like to thank Andrea Addazi, Marcus Berg, Marco Bochicchio, Dario Consoli, Giuseppe D'Appollonio, Paolo Di Vecchia, Yu-tin Huang, Raffaele Marotta, Francisco Morales, Tassos Petkou, Lorenzo Pieri, Oliver Schlotterer, Stephan Stieberger, Gabriele Veneziano, and Congkao Wen, for interesting discussions. This work is partially supported by the INFN network ST\&FI {\it ``String Theory and Fundamental Interactions''}.

\appendix
\section{Expansion of the amplitudes}

\subsection{Soft limit of the amplitudes with the Konishi operator}
In this section we give more details about the soft limit of the amplitudes in Eqs.~\eqref{MGGGK}, \eqref{MDDGK} and \eqref{MDGGK}.
As a preliminary step we consider the soft limit of the common dynamical factor when the momentum $k_3$ becomes soft in any case.
\be
\frac{\Gamma\left(k_3p_4\right)\Gamma\left(1+ k_2k_3\right)\Gamma\left(1+ k_1 k_3 \right)}{\Gamma\left(2- k_3p_4\right)\Gamma\left(- k_2k_3\right)\Gamma\left(1- k_1k_3\right)}=\frac{k_2k_3}{ k_3p_4 k_1k_2}+\cO(\delta^3)=\frac{\lef 23 \re [32]}{\lef 12 \re [21] k_3p_4}+\cO(\delta^3),
\label{DynKonishi}
\ee 
Combining this expression with the expansions of the different kinetic terms we will get the final result.

\subsubsection{The amplitude $\cM_4(\cE_1^{-2},\cE_2^{+2},\cE_3^{+2},\cK_4^{+4})$}
\label{appendixGGGK}
The expansion of the kinematical term in Eq.~\eqref{MGGGK} yields
\begin{align}
&G_N\frac{[13]\lef 14 \re^8}{\lef 12 \re \lef 23 \re m_H^6}\frac{[12] [45]^4}{\lef 13 \re \lef 32 \re}\nonumber\\
&\qquad\qquad\qquad{=}G_N\frac{[13][12]\lef 14\re^4}{\lef 12 \re \lef 13 \re \lef 23 \re \lef 32 \re m^6_H}\lef 12 \re^4 [25]^4\left(1{+}4\delta\frac{\lef 13 \re [35]}{\lef 12 \re [25]}{+}6\delta^2\frac{\lef 13 \re^2 [35]^2}{\lef 12 \re^2 [25]^2}\right){+}\cO(\delta^3).
\label{kinMGGGK}
\end{align}
Combining the Eq.~\eqref{DynKonishi} with Eq.~\eqref{kinMGGGK} we obtain up to order $\delta$ terms
\begin{align}
\cO(\delta^{-1}):\quad\quad &G_N\frac{[13][23]\lef 12 \re^2}{\lef 13 \re \lef 32 \re 2k_3p_4}\frac{\lef 14 \re^4[25]^4}{m^6}\\
\cO(\delta^{0}):\quad\quad &4G_N\frac{[13][23][35][25]^3\lef 12 \re \lef 14 \re^4}{\lef 32 \re  2k_3p_4 m^6}\\
\cO(\delta):\quad\quad &6G_N\frac{[13][23][35]^2[25]^2\lef 13 \re\lef 14 \re^4}{\lef 32 \re 2k_3p_4 m^6}.
\end{align}
\subsubsection{The amplitude $\cM_4(\phi_1,\phi_2,\cE_3^{+2},\cK_4^{+4})$}
\label{appendixDDGK}
The kinematical term in Eq.~\eqref{MDDGK} yields
\begin{align}
G_N\frac{[13][12]\lef 14 \re^4\lef 24 \re^4 [45]^4}{\lef 12 \re \lef 23 \re \lef 13 \re \lef 32 \re m_H^6}&{=}G_N\frac{[12][13]\lef 14\re^2\lef 24\re^2}{\lef 12 \re \lef 23 \re \lef 13 \re \lef 32 \re m_H^6}\lef 12 \re^4[25]^2[15]^2\nonumber\\
&\left(1+2\delta\left( \frac{\lef 13 \re[35]}{\lef 12 \re [25]}+\frac{\lef 23 \re[35]}{\lef 12 \re [15]} \right)+\delta^2\left( \frac{\lef 13 \re^2 [35]^2}{\lef 12 \re^2 [25]^2}+\frac{\lef 23 \re^2 [35]^2}{\lef 12 \re^2 [15]^2} \right)\right).
\end{align}
Here we give the result of the expansion to be compared with the predictions dictated by the soft theorem.
\begin{align}
\cO(\delta^{-1}):\quad\quad &G_N\frac{\lef 12 \re^2[13][23]}{ \lef 13 \re \lef 32 \re 2k_3p_4}\frac{\lef 14\re^2 \lef 24\re^2 [25]^2[15]^2}{m_H^6}\\
\cO(\delta^{0}):\quad\quad &2G_N\frac{\lef 12 \re [13][23]}{ \lef 13 \re \lef 32 \re 2k_3p_4}\frac{\lef 14\re^2 \lef 24\re^2 [25]^2[15]^2}{m_H^6}\left(\frac{\lef 13 \re [35]}{[25]}+\frac{\lef 23 \re [25]}{[15]}\right)\\
\cO(\delta):\quad\quad &G_N\frac{[13][23]}{ \lef 13 \re \lef 32 \re 2k_3p_4}\frac{\lef 14\re^2 \lef 24\re^2 [25]^2[15]^2}{m_H^6}\left(\frac{\lef 13 \re^2 [35]^2}{[25]^2}+\frac{\lef 23^2 \re [25]^2}{[15]^2}\right).
\end{align}

\subsubsection{The amplitude $\cM_4(\phi_1,\cE_2^{+2},\cE_3^{+2},\cH_4^{+2})$}
\label{appendixDGGK}
To expand the amplitude in Eq.~\eqref{MDGGK} we need to expand only $\cA_L$
\be
G_N\frac{[13][12]\lef 14\re^4[45]^2}{\lef 12 \re \lef 13 \re \lef 23 \re \lef 32\re}=G_N\frac{[13][12]\lef 14 \re^2[25]^2 \lef 12 \re}{\lef 13 \re \lef 23 \re \lef 32 \re}\left( 1+2\delta\frac{\lef 13\re [35]}{\lef 12\re [25]}+\delta^2\frac{\lef 13\re^2 [35]^2}{\lef 12\re^2 [25]^2} \right), 
\ee
getting
\begin{align}
\cO(\delta^{-1}):\quad\quad &G_N\frac{[13][23]\lef 12 \re^2}{\lef 13 \re \lef 32 \re 2k_3p_4}\frac{\lef 14 \re^2[25]^2 [12]^2}{m^4}\\
\cO(\delta^{0}):\quad\quad &2G_N\frac{[13][23]\lef 12 \re}{\lef 32 \re 2k_3p_4}\frac{\lef 14 \re^2[25]  [35] [12]^2}{m^4}\\
\cO(\delta):\quad\quad &G_N\frac{[13][23]}{\lef 32 \re 2k_3p_4}\frac{\lef 13 \re \lef 14 \re^2 [35]^2 [12]^2}{m^4}.
\end{align}

\subsection{The amplitude $\cM_4(\cE_1,\cE_2,\Tac_3,\Tac_4)$}
\label{appendixTTET}
It is convenient to factor out the structure in Eq.~\eqref{Gamma3}, which has a trivial soft behavior, from the dynamical term in Eq.~\eqref{dynMTTET}
\be
\cI(s,t,u)=-\frac{k_1p_3\,k_1p_4}{(1-k_1k_2)^2}(1{+}\cO(\delta^3))= -k_1k_2\,k_1p_3\,k_1p_4\left( \frac{1}{k_1k_2} {+} 2 {+} 3 k_1k_2\right){+}\cO(\delta^3).
\ee 
The expansion of the kinematical structure $\cE_1 \cK \cE_2^t \cK^t$ can be organized as follows
\be
\cE_1 \cK \cE_2^t \cK_2^t=\cE_1 \cK_{-1} \cE_2^t \cK_{-1}^t{+}2\cE_1 \cK_0 \cE_2^t \cK_{-1}^t{+}\cE_1 \cK_0 \cE_2^t \cK_0^t,
\ee
where
\begin{align}
\cK_{-1}&=\frac{p_3 \otimes p_4}{k_1p_3}{+}\frac{p_4\otimes p_3}{k_1p_4}\\
\cK_0&=1-p_3\otimes p_3-p_4\otimes p_4 {+} \frac{p_3\otimes p_4}{k_1p_3}k_1p_4{+}\frac{p_4\otimes p_3}{k_1p_4}k_1p_3.
\end{align}
The expansion of the amplitude up to $\cO(\delta)$ yields
\begin{align}
&\cM(\cE_1,\cE_2,\Tac_3,\Tac_4)={-}\frac{1}{\delta}\left(\frac{k_1p_3\,k_1p_4}{k_1k_2}\cE_1 \cK_{-1} \cE_2^t \cK_{-1}^t\right)\nonumber\\
&+\delta^0\left({-}2\frac{k_1p_3\,k_1p_4}{k_1k_2} \cE_1 \cK_0 \cE_2^t \cK_{-1}^t{-}2k_1p_3\,k_1p_4 \,\cE_1 \cK_{-1} \cE_2^t \cK_{-1}^t\right)\nonumber\\
&+\delta\left({-}\frac{k_1p_3\,k_1p_4}{k_1k_2} \cE_1 \cK_0 \cE_2^t \cK_0^t {-} 4\,k_1p_3\,k_1p_4 \, \cE_1 \cK_0 \cE_2^t \cK_{-1}^t {-} 3\,k_1k_2\,k_1p_3\,k_1p_4 \, \cE_1 \cK_{-1} \cE_2^t \cK_{-1}^t{{+}}\right)\cO(\delta^2).
\end{align}
To make explicitly the expansion it is convenient to introduce the variables $p_{+}=p_3{+}p_4$ and $p_-=p_3-p_4$. As far as $\cE_2$ is concerned, all the bilinear 
expressions involving $\cE_2$ are well organized
\begin{align}
p_-\cE_2p_-{=}\cO(1);\qquad
p_{+}\cE_2p_-{=}-k_1\cE_2p_-{=}\cO(\delta);\qquad
p_{+}\cE_2p_{+}{=}k_1\cE_2k_1{=}\cO(\delta^2).
\end{align}

Starting with the tensorial structure $\cE_1 \cK_{{-}1} \cE_2^t \cK_{{-}1}^t$ respectively for $\cE_{1,2} = h/\phi$ both symmetric (graviton and dilaton) and for  $\cE_{1,2} = b$ both antisymmetric (Kalb{-}Ramond fields) we get up to $\cO(1)$
\begin{align}
&\cE_1 \cK_{{-}1} \cE_2^t \cK_{{-}1}^t(\delta^{{-}2}) =\frac{1}{4}\left( \frac{p_3\cE_1p_3}{(k_1p_3)^2}{+}\frac{p_4\cE_1p_4}{(k_1p_4)^2}{-}2\frac{p_3\cE_1p_4}{k_1p_3\,k_1p_4}\right) p_{-}\cE_2p_{-}\nonumber\\
&\cE_1 \cK_{{-}1} \cE_2^t \cK_{{-}1}^t(\delta^{{-}1})=\frac{1}{2}\left({-}\frac{p_3\cE_1p_3}{(k_1p_3)^2}{+}\frac{p_4\cE_1p_4}{(k_1p_4)^2}\right) p_{+}\cE_2p_{-}\nonumber\\
&\cE_1 \cK_{{-}1} \cE_2^t \cK_{{-}1}^t(\delta^0)=\frac{1}{4}\left( \frac{p_3\cE_1p_3}{(k_1p_3)^2}{+}\frac{p_4\cE_1p_4}{(k_1p_4)^2}{+}2\frac{p_3\cE_1p_4}{k_1p_3\,k_1p_4}\right) p_{+}\cE_2p_{+}.
\end{align}

\begin{align}
b_1 \cK_{{-}1} b_2^t \cK_{{-}1}^t=\frac{p_3b_1p_4}{k_1p_3k_1p_4}p_{+}b_2p_{-}.
\end{align}

The expansion of the structure $2\cE_1 \cK_0 \cE_2^t \cK_{{-}1}^t$ is up to $\cO(1)$
\begin{align}
2\cE_1 \cK_0 \cE_2^t \cK_{{-}1}^t(\delta^{{-}1})&=
\frac{2}{k_1p_3}\left({-}p_3\cE_1\cE_2\frac{p_{-}}{2}{{+}}\frac{1}{4}p_{-}\cE_2p_{-}\left(p_3\cE_1p_{-}{{+}}p_3\cE_1p_3\frac{k_1p_4}{k_1p_3}{{-}}p_3\cE_1p_4\frac{k_1p_3}{k_1p_4}\right)\right)\nonumber\\
&{+}\frac{2}{k_1p_4}\left(p_4\cE_1\cE_2\frac{p_{-}}{2}{+}\frac{1}{4}p_{-}\cE_2p_{-}\left({-}p_4\cE_1p_{-}{-}p_4\cE_1p_3\frac{k_1p_4}{k_1p_3}{+}p_4\cE_1p_4\frac{k_1p_3}{k_1p_4}\right)\right)\nonumber\\
2\cE_1 \cK_0 \cE_2^t \cK_{{-}1}^t(\delta^{0})&=
\frac{2}{k_1p_3}\left(p_3\cE_1\cE_2\frac{p_{+}}{2}{+}\frac{1}{2}p_{+}\cE_2p_{-}\left(p_3\cE_1p_4 {-}p_3\cE_1p_3\frac{k_1p_4}{k_1p_3}\right)\right)\nonumber\\
&{+}\frac{2}{k_1p_4}\left(p_4\cE_1\cE_2\frac{p_{+}}{2}{+}\frac{1}{2}p_{+}\cE_2p_{-}\left({-}p_4\cE_1p_3 {{+}} p_4\cE_1p_4\frac{k_1p_3}{k_1p_4}\right)\right).
\end{align}
\begin{align}
&2b_1 \cK_0 b_2^t \cK_{{-}1}^t=\frac{p_3b_1b_2p_{-}}{k_1p_3}{-}\frac{p_4b_1b_2p_{-}}{k_1p_4}\nonumber\\
&{-}\frac{p_3b_1b_2p_{+}}{k_1p_3}{-}\frac{p_4b_1b_2p_{+}}{k_1p_4}{+}\left(\frac{p_3b_1p_4}{k_1p_4}{-}\frac{p_4b_1p_3}{k_1p_3}\right)p_{+}b_2p_{-}.
\end{align}

Finally we consider the expansion of the structure $\cE_1 \cK_0 \cE_2^t \cK_0^t$
\begin{align}
&\cE_1 \cK_0 \cE_2^t \cK_0^t(\delta^0)=\cE_1\cE_2{-}\frac{1}{2}p_{-}\{\cE_1,\cE_2\}p_{-}{-}\frac{k_1p_4}{k_1p_3}p_3\cE_1\cE_2p_{-}{+}\frac{k_1p_3}{k_1p_4}p_4\cE_1\cE_2p_{-}\nonumber\\
&{+}\frac{1}{2}p_{-}\cE_2p_{-}\biggr(\frac{1}{2}p_3\cE_1p_3{+}\frac{1}{2}p_4\cE_1p_4{-}p_3\cE_1p_4{+}p_3\cE_1p_3\frac{k_1p_3}{k_1p_4}{+}p_4\cE_1p_4\frac{k_1p_3}{k_1p_4}\nonumber\\&{-}p_3\cE_1p_4\frac{k_1p_3}{k_1p_4}{-}p_3\cE_1p_4\frac{k_1p_3}{k_1p_4}{+}\frac{1}{2}p_3\cE_1p_3\left(\frac{k_1p_4}{k_1p_3}\right)^2{+}\frac{1}{2}p_4\cE_1p_4\left(\frac{k_1p_3}{k_1p_4}\right)^2\biggr).
\end{align}
\begin{align}
b_1 \cK_0 b_2^t \cK_0^t={-}b_1b_2{+}\frac{1}{2}p_{-}\{b_1,b_2\}p_{-} {+}\frac{k_1p_4}{k_1p_3}p_3b_1b_2p_{-}{-}\frac{k_1p_3}{k_1p_4}p_4b_1b_2p_{-}.
\end{align}

Now we have all the ingredients to compute the full expansion of the amplitude.
Consider first the symmetric case in which $\cE_{1/2}=h/\phi$.
At leading order we have
\begin{align}
&{-}\frac{k_1p_3\,k_1p_4}{k_1k_2}\cE_1 \cK_{{-}1} \cE_2^t \cK_{{-}1}^t={-}\frac{1}{4}\left(p_3\cE_1p_3 \frac{k_1p_4}{k_1k_2\,k_1p_3}{+}p_4\cE_1p_4 \frac{k_1p_3}{k_1k_2\,k_1p_4} {-} 2\frac{p_3\cE_1p_4}{k_1k_2} \right)p_{-}\cE_2p_{-}\nonumber\\
&=\left( \frac{p_3\cE_1p_3}{k_1p_3} {+}\frac{p_4\cE_1p_4}{k_1p_4}{+}\frac{k_2\cE_1k_2}{k_1k_2}\right)\frac{p_{-}}{2}\cE_2\frac{p_{-}}{2},
\end{align}
which has the expected structure from the soft theorem.

The subleading order comes from three different contributions: 
\begin{align}
&\cI_1 \times 2\cE_1 \cK_0 \cE_2^t \cK_{{-}1}^t={-}2\frac{k_1p_4}{k_1k_2}\biggr( {-}p_3\cE_1\cE_2\frac{p_{-}}{2}{+}\frac{1}{4}p_{-}\cE_2p_{-}\biggr(p_3\cE_1p_{-}{+}p_3\cE_1p_3\frac{k_1p_4}{k_1p_3}{-}p_3\cE_1p_4 \frac{k_1p_3}{k_1p_4}\biggr)\biggr)\nonumber\\
&{-}2\frac{k_1p_3}{k_1k_2}\biggr( p_4\cE_1\cE_2\frac{p_{-}}{2}{+}\frac{1}{4}p_{-}\cE_2p_{-}\biggr({-}p_4\cE_1p_{-}{-}p_4\cE_1p_3\frac{k_1p_4}{k_1p_3}{+}p_4\cE_1p_4 \frac{k_1p_3}{k_1p_4}\biggr)\biggr).
\end{align}
\begin{align}
\cI_{2}\times \cE_1 \cK_{{-}1} \cE_2^t \cK_{{-}1}^t={-}\frac{1}{2}\left( p_3\cE_1p_3\frac{k_1p_4}{k_1p_3}{-}2p_3\cE_1p_4{+}p_4\cE_1p_4\frac{k_1p_3}{k_1p_4}\right)p_{-}\cE_2p_{-}
\end{align}

The subleading contribution coming from
\begin{align}
\cI_1\times \cE_1 \cK_{{-}1} \cE_2^t \cK_{{-}1}^t={-}\frac{1}{2}\left( {-}p_3\cE_1p_3\frac{k_1p_4}{k_1k_2\,k_1p_3}{+}p_4\cE_1p_4\frac{k_1p_3}{k_1k_2\,k_1p_4}\right)p_{+}\cE_2p_{-}.
\end{align}

The sum of these three gives the answer expected from the soft graviton theorem
\begin{align}
&\frac{1}{2k_1p_3}p_3\cE_1p_3\,p_{+}\cE_2p_{-}{+}\frac{1}{2}\left(p_3\cE_1p_3{-}p_4\cE_1p_4\right)\frac{p_{+}\cE_2p_{-}}{k_1k_2}{-}\frac{1}{2k_1p_4}p_4\cE_1p_4\,p_{+}\cE_2p_{-}\nonumber\\
&{-}p_3\cE_1\cE_2p_{-}\frac{k_1p_4}{k_1k_2}{+}p_4\cE_1\cE_2p_{-}\frac{k_1p_3}{k_1k_2}\nonumber\\
&=\frac{p_3\cE_1p_3}{2k_1p_3}p_{+}\cE_2p_{-} {+}\frac{1}{2}p_3\cE_1\cE_2p_{-} {-} \frac{p_4\cE_1p_4}{2k_1p_4}p_{+}\cE_2p_{-} {-} \frac{1}{2}p_4\cE_1\cE_2p_{-}\nonumber\\
&{-}\frac{k_1p_{-}}{2k_1k_2}k_2\cE_1\cE_2p_{-}{-}\frac{k_2\cE_1p_{-}}{2k_1k_2}p_{+}\cE_2p_{-}.
\end{align}
It is straightforward to compare the last expression with the expected behavior
\begin{align}
\frac{k_1J_2\cE_1k_2}{k_1k_2}\frac{p_{-}}{2}\cE_2\frac{p_{-}}{2}&=\frac{k_2\cE_1p_{-}}{2k_1k_2}k_1\cE_2p_{-}{-}\frac{k_1p_{-}}{2k_1k_2}k_2\cE_1\cE_2p_{-}\\
\frac{k_1J_3\cE_1p_3}{k_1p_3}\frac{p_{-}}{2}\cE_2\frac{p_{-}}{2}&=p_3\cE_1\cE_2p_{-}\frac{k_1p_3}{2k_1p_3}{-}\frac{p_3\cE_1p_3}{2k_1p_3}k_1\cE_2p_{-} \\
\frac{k_1J_4\cE_1p_4}{k_1p_4}\frac{p_{-}}{2}\cE_2\frac{p_{-}}{2}&= {-} p_4\cE_1\cE_2p_{-}\frac{k_1p_4}{2k_1p_4}{+} \frac{p_4\cE_1p_4}{2k_1p_4}k_1\cE_2p_{-} .
\end{align}

The sub{-}sub{-}leading contribution comes from the sum of the following terms
\begin{align}
\cM_1={-}\frac{k_1p_3\,k_1p_4}{k_1k_2} \cE_1 \cK_0 \cE_2^t \cK_0^t(\delta^0) {-} 4\,k_1p_3\,k_1p_4 \, \cE_1 \cK_0 \cE_2^t \cK_{{-}1}^t(\delta^{{-}1}) \nonumber\\
{-} 3\,k_1k_2\,k_1p_3\,k_1p_4 \, \cE_1 \cK_{{-}1} \cE_2^t \cK_{{-}1}^t(\delta^{{-}2}){-}2\frac{k_1p_3\,k_1p_4}{k_1k_2}\, \cE_1 \cK_0 \cE_2^t \cK_{{-}1}^t(\delta^{0})\nonumber\\
{-}2\,k_1p_3\,k_1p_4 \, \cE_1 \cK_{{-}1} \cE_2^t \cK_{{-}1}^t(\delta^{{-}1}){-}\frac{k_1p_3\,k_1p_4}{k_1k_2} \cE_1 \cK_{{-}1} \cE_2^t \cK_{{-}1}^t(\delta^0).
\label{subsubEETT}
\end{align}
%

In Eq.~\eqref{subsubEETT} we can recognize the structures predicted by the soft theorem
\begin{align}
\frac{1}{2}\frac{k_1J_3\cE_1J_3k_1}{k_1p_3}=\frac{1}{2}p_3\cE_1\cE_2k_1{-}\frac{1}{4k_1p_3}k_1\cE_2k_1\,p_3\cE_2p_3{-}\frac{1}{4}k_1p_3\,\cE_1\cE_2\\
\frac{1}{2}\frac{k_1J_4\cE_1J_4k_1}{k_1p_4}=\frac{1}{2}p_4\cE_1\cE_2k_1{-}\frac{1}{4k_1p_4}k_1\cE_2k_1\,p_4\cE_2p_4{-}\frac{1}{4}k_1p_4\,\cE_1\cE_2\\
\frac{1}{2}\frac{k_1J_4\cE_1J_4k_1}{k_1p_4}=k_1p_{-}\frac{ p_{-}\cE_1\cE_2k_1}{2k_1k_2}{-}\frac{p_{-}\cE_1p_{-} ~k_1\cE_2k_1}{4k_1k_2}{-}\frac{\cE_1\cE_2 (k_1p_{-})^2}{4k_1k_2}
\end{align}
and additional terms
\be
\alch\left({-}\frac{k_1p_{-}}{2}k_2\cE_1\cE_2p_{-}{+}\frac{k_1k_2}{2}p_{-}\cE_1\cE_2p_{-}{+}\frac{k_1p_{-} ~\ k_1\cE_2p_{-}~\ k_2\cE_1p_{-}}{2k_1k_2}{-}\frac{k_1\cE_2p_{-}~\ k_2\cE_1p_{-}}{2}\right).
\label{TTETadditional}
\ee
    
For two Kalb{-}Ramond fields $\cE^t_{1,2} = {-}  \cE^t_{1,2} $ the amplitude at order $\cO(\delta^{{-}1})$ is zero. 
The expansion starts at order $\cO(\delta^0)$ with 
\begin{align}
\cM_0=\frac{p_3b_1p_4}{k_1k_2}p_{+}b_2p_{-}{-}\frac{k_1p_{-}}{2k_1k_2}p_{+}b_1b_2p_{-}{-}\frac{1}{4}p_{-}\{b_1,b_2\}p_{-}.
\end{align}
Note that $p_{+}b_2 = {-} k_1b_2$ is of order $\cO(\delta)$, while $p_{+}b_1 = {-} k_2b_1$ is of order $\cO(\delta^0)$.
At order $\cO(\delta)$ the amplitude looks like
\begin{align}
\cM_1=p_3b_1p_4\,p_{+}b_2p_{-}{-}b_1b_2\frac{k_1p_4\,k_1p_3}{k_1k_2}{-}\frac{1}{2}p_{+}b_1b_2p_{-}\,k_1p_{-}\nonumber\\
{-}\frac{1}{4}p_{-}\{ b_1,b_2\}p_{-}\,k_1k_2{+}\frac{1}{4}p_{+}\{b_1,b_2\}p_{+}{+}\frac{k_1p_{-}}{2k_1k_2}p_{-}b_1b_2p_{+}.
\end{align}

 
\subsection{The amplitude $\cM_4(\Tac_1,\Tac_2,\cE_3,\cK_4)$}
\label{appendixTTEK}

The expansion is organized as follows.
The dynamical term can be expanded as
\be
 \frac{\Gamma(1{+} p_1k_3)\Gamma({-}1{+} k_3p_4)\Gamma(1{+} p_2k_3)}{\Gamma({-} p_1k_3) \Gamma(2{-} k_3p_4) \Gamma({-} p_2k_3)}={-}\frac{k_3p_1\,k_3p_2}{k_3p_4}(1{+}2k_3p_4{+}3(k_3p_4)^2){+}\mathcal{O}(\delta^3).
\ee

The expansion of the open string amplitude $\Amp_L$ can be expanded up to the $\mathcal{O}(\delta^0)$ as
\begin{align}
&\Amp^{({-}1)}_L=\frac{1}{\delta} \left( \frac{a_3p_2}{k_3p_2}{-}\frac{a_3p_1}{k_3p_1} \right) p_2Hp_2 \nonumber\\
&\Amp^{(0)}_L =\delta^0 \left[ \left( {-} \frac{a_3p_2}{k_3p_2}k_3p_4{+}\frac{a_3p_1}{k_3p_1}k_3p_4 \right) {-}2a_3Hp_1{+}2a_3p_4 \,p_2Hp_2{-}2a_3p_2\frac{p_2Hk_3}{k_3p_2} \right].
\end{align}
The expansion for $\Amp_R$ is obtained by exchanging the labels $1 \leftrightarrow 2$.

The expansion of the kinematical term of the amplitude $\cM(\Tac_1,\Tac_2,\cE_3,\cK_4)$ can be easily yield multiplying the expansions of the open string amplitudes.
\begin{align}
K^{({-}2)}&=\Amp_L^{({-}1)}\otimes \Amp_R^{({-}1)}\nonumber\\
K^{({-}1)}&=\Amp_L^{({-}1)}\otimes \Amp_R^{(0)}{+}\Amp_L^{(0)}\otimes \Amp_R^{({-}1)}\nonumber\\
K^{(0)}&=\Amp_L^{(0)}\otimes \Amp_R^{(0)}.
\end{align}

For simplicity we consider only the case in which $cK_4$ is the completely symmetric irreducible state.
To complete the expansion we need to disentangle the sub{-}leading contributions to each $\cK^{(i)}$ term.
\begin{align}
&\cK^{({-}2)}(\delta^{{-}2})=\left( {-}\frac{p_1\cE_3p_1}{k_3p_1}{-}\frac{p_2\cE_3p_2}{k_3p_2} {+} 2\frac{p_1\cE_3p_2}{k_3p_1 k_3p_2}\right)\cK_4\left[\pmm,\pmm,\pmm,\pmm\right]\\
&\cK^{({-}2)}(\delta^{{-}1})=0\nonumber\\
&\cK^{({-}2)}(\delta^{0})=\left( \frac{p_1\cE_3p_1}{2(k_3p_1)^2}{-}\frac{p_1\cE_3p_2}{k_3p_1\,k_3p_2}{+}\frac{p_2\cE_3p_2}{2(k_3p_2)^2} \right)\cK_4\left[\pmm,\pmm,k_3,k_3\right].
\end{align}

\begin{align}
\cK^{({-}1)}(\delta^{{-}1})&=2\left( {-}\frac{p_1\cE_3p_2}{(k_3p_1)^2}{+}\frac{p_2\cE_3p_2}{(k_3p_2)^2}\right)\cK_4\left[\pmm,\pmm,\pmm,k_3\right]\nonumber\\
&{+}2\left( \frac{p_1\cE_3p_1}{(k_3p_1)^2}{-}2\frac{p_1\cE_3p_2}{k_3p_1\,k_3p_2}{+}\frac{p_2\cE_3p_2}{(k_3p_2)^2} \right)k_3p_4\,\cK_4\left[\pmm,\pmm,\pmm,\pmm\right]\nonumber\\
&{+}\frac{(p_1\cE_3)^\mu}{2k_3p_1}\cK_4\left[ \mu,\pmm,\pmm,\pmm \right] {-}\frac{(p_2\cE_3)^\mu}{2k_3p_2}\cK_4\left[ \mu,\pmm,\pmm,\pmm \right],\nonumber\\
\cK^{({-}1)}(\delta^{0})&=2\left( \frac{p_1\cE_3p_1}{(k_3p_1)^2}{+}\frac{p_2\cE_3p_2}{(k_3p_2)^2}{-}2\frac{p_1\cE_3p_2}{k_3p_1\,k_3p_2}\right)\pmm H k_3 \pmm H k_3
\end{align}

\begin{align}
\cK^{(0)}(\delta^{0})&={-}\cE_3^{\mu\nu}\cK_4\left[\mu,\nu,\pmm,\pmm\right]{-}\frac{p_1\cE_3p_2}{k_3p_1\,k_3p_2}\cK_4\left[\pmm,\pmm,k_3,k_3 \right]\nonumber\\
&{+}2\left( \frac{p_1\cE_3p_1}{(k_3p_1)^2}k_3p_4{+}2\frac{p_1\cE_3p_4}{k_3p_1}{-}\frac{p_2\cE_3p_2}{(k_3p_2)^2}k_3p_4{-}2\frac{p_2\cE_3p_4}{k_3p_2} \right)\cK_4\left[\pmm,\pmm,\pmm,k3\right]\nonumber\\
&{+}\left( {-}\frac{p_1\cE_3p_1}{(k_3p_1)^2}(k_3p_4)^2{+}2\frac{p_1\cE_3p_2}{k_3p_1\,k_3p_2}(k_3p_4)^2{-}\frac{p_2\cE_3p_2}{(k_3p_2)^2}(k_3p_4)^2 \right)\cK_4\left[ \pmm,\pmm,\pmm,\pmm\right]\nonumber\\
&{+}\frac{1}{4}p_4\cE_3p_4\,\cK_4\left[ \pmm,\pmm,\pmm,\pmm \right]{+}\left(\frac{(p_1\cE_3)^\mu}{k_3p_1}{+}\frac{(p_2\cE_3)^\mu}{k_3p_2}\right)\cK_4\left[\mu,\pmm,\pmm,k_3\right]\nonumber\\
&{+}\left( {-}\frac{(p_1\cE_3)^\mu}{2k_3p_1}{+}\frac{(p_2\cE_3)^\mu}{2k_3p_2} \right)k_3p_4\,\cK_4\left[ \mu,\pmm,\pmm,\pmm \right].
\end{align}

Collecting the contributions to each term in the soft expansion, we reproduce the leading and sub{-}leading behavior as predicted by the soft theorem, but we find that 
the sub{-}sub{-}leading order is not.
\begin{align}
\cM^{({-}1)}=\frac{k_3p_1\,k_3p_2}{k_3p_4}\cK^{({-}2)}(\delta^{{-}2})=\left(\frac{p_1\cE_3p_1}{k_3p_1}{+}\frac{p_2\cE_3p_2}{k_3p_2}{+}\frac{p_4\cE_3p_4}{k_3p_4}\right)\cK_4\left[ \pmm,\pmm,\pmm,\pmm \right]\end{align}

\begin{align}
\cM^{(0)}&=\frac{k_3p_1\,k_3p_2}{k_3p_4}\cK^{({-}2)}(\delta^{{-}1}){+}\frac{k_3p_1\,k_3p_2}{k_3p_4}\cK^{({-}1)}(\delta^{{-}1}){+}2k_3p_1\,k_3p_2\cK^{({-}2)}(\delta^{{-}2})\nonumber\\
&=2\left(\frac{p_1\cE_3p_1}{k_3p_1}{-}\frac{p_2\cE_3p_2}{k_3p_2}{-}\frac{p_{-}\cE_3p_4}{k_3p_4}\right)\cK_4\left[\pmm,\pmm,\pmm,k_3\right]\nonumber\\
&{+}2\left({-}(p_1\cE_3)^\mu{+}(p_2\cE_3)^\mu{+}\frac{(p_4\cE_3)^\mu}{k_3p_4}k_3p_{-}\right)\cK_4\left[ \mu,\pmm,\pmm,\pmm \right].
\end{align}

For the sake of completeness we report the expression obtained at the sub-sub-leading order.
\begin{align}
\cM^{(1)}&=-\frac{k_3p_1\,k_3p_2\,\cK_4[\cE_3,\pmm,\pmm]}{k_3p_4}+\frac{k_3p_1\,k_3p_2\,p_4\cE_3p_4\,\cK_4[\pmm,\pmm,\pmm,\pmm]}{4k_3p_4}\nonumber\\
&+\left( \frac{p_1\cE_3p_1}{8k_3p_1}{+}\frac{p_1\cE_3p_1}{8k_3p_4}{-}3\frac{p_1\cE_3p_2}{4k_3p_4}{-}\frac{p_2\cE_3p_2}{8k_3p_2}{+}\frac{p_2\cE_3p_2}{8k_3p_4} \right)\cK_4[\pmm,\pmm,k_3,k_3]\nonumber\\
&+\left(-\frac{k_3p_2\,p_1\cE_3p_1}{4k_3p_1}+\frac{k_3p_2\,p_1\cE_3p_4}{2k_3p_4}+\frac{k_3p_1\,p_2\cE_3p_2}{4k_3p_2}-\frac{k_3p_1\,p_2\cE_3p_4}{2k_3p_4}\right)\cK_4[\pmm,\pmm,\pmm,k_3]\nonumber\\
&+\left( \frac{k_3p_2\,(p_1\cE_3)^\mu}{k_3p_4} +\frac{k_3p_1\,(p_2\cE_3)^\mu}{k_3p_4}\right)\cK_4[_\mu,\pmm,\pmm,k_3]\nonumber\\
&+\frac{1}{2}\left( k_3p_2\,(p_1\cE_3)^\mu-k_3p_1\,(p_2\cE_3)^\mu \right)\cK_4[_\mu,\pmm,\pmm,\pmm].
\end{align}

\end{document}